\documentclass{article}

\usepackage{arxiv}

\usepackage[utf8]{inputenc} 
\usepackage[T1]{fontenc}    
\usepackage{url}            
\usepackage{booktabs}       
\usepackage{nicefrac}       
\usepackage{microtype}      
\usepackage{lipsum}		
\usepackage{graphicx}
\usepackage{natbib}
\usepackage{doi}
\usepackage{amsmath,amsfonts,amssymb}
\usepackage{setspace}
\usepackage{tocloft}
\usepackage{float}
\usepackage[table]{xcolor}
\usepackage{macro_ref}
\usepackage{caption}
\usepackage{subcaption}
\usepackage{lineno}
\usepackage{soul}

\title{Asgard/NOTT: First lab assembly and experimental results}

\author{}

\date{}

\renewcommand{\shorttitle}{Garreau et al.: Asgard/NOTT: First lab assembly and experimental results}

\hypersetup{
pdftitle={Asgard/NOTT: First lab assembly and experimental results},
pdfsubject={astro-ph.IM},
pdfauthor={Germain Garreau},
pdfkeywords={instrumentation; integrated optics; nulling interferometry; Very Large Telescope Interferometer; Asgard/NOTT},
}

\begin{document}
\maketitle
\vspace{-6em}
\begin{center}
G. Garreau$^{a}$\footnote{germain.garreau@kuleuven.be},
A. Bigioli$^{a}$,
R. Laugier$^{a}$,
B. La Torre$^{b}$,
M-A. Martinod$^{a}$,
K. Missiaen$^{a}$,
J. Morren$^{a}$,
G. Raskin$^{a}$,
M. Salman$^{a}$,
S. Gross$^{c}$,
M. Ireland$^{d}$,
A. P. Joó$^{e}$,
L. Labadie$^{f}$,
S. Madden$^{d}$,
A. Mazzoli$^{g}$,
G. Medgyesi$^{e}$,
A. Sanny$^{c,f}$,
A. Taras$^{h}$,
B. Vandenbussche$^{a}$, and
D. Defrère$^{a}$
~\newline

$^a$Institute of Astronomy, KU Leuven, Celestijnenlaan 200D, 3001 Leuven, Belgium\\
$^b$Institut d'Optique Graduate School, Université Paris-Saclay, 91127 Palaiseau cedex, France\\
$^c$MQ Photonics Research Centre, School of Mathematical and Physical Sciences, Macquarie University, NSW, 2109, Australia\\
$^d$Research School of Astronomy \& Astrophysics, Australian National University, Canberra, ACT 2611, Australia\\
$^e$Konkoly Observatory, 	HUN-REN Research Centre for Astronomy and Earth Sciences, Budapest, Hungary\\
$^f$I. Physikalisches Institut, Universität zu Köln, Zülpicher Str. 77, 50937 Köln, Germany\\
$^g$STAR Institute, University of Liège, 19C allée du Six Août, 4000 Liège, Belgium\\
$^h$Astralis-Usyd, Sydney Institute for Astronomy, School of Physics, University of Sydney, NSW 2006, Australia
\end{center}
~\vspace{-0.5em}

\begin{abstract}
    Asgard/NOTT is an ERC-funded project hosted at KU Leuven and is part of a new visitor instrumental suite, called Asgard, under preparation for the Very Large Telescope Interferometer (VLTI). Leveraging nulling capabilities and the long VLTI baselines, it is optimized for high-contrast imaging of the snow line region around young nearby main-sequence stars. This will enable the characterization of the atmosphere of young giant exoplanets and warm/hot exozodiacal dust with spectroscopy in the L’-band (3.5-4.0\,µm). In this work, we present the first lab assembly of the instrument done at KU Leuven and the technical solutions to tackle the challenge of performing nulling in the mid-infrared despite the thermal background. The opto-mechanical design of the warm optics and the injection system for the photonic chip are described. The alignment procedure used to assemble the system is also presented. Finally, the first experimental results, including fringes and null measurements, are given and confirm the adequacy of the bench to test and optimize the Asgard/NOTT instrument. 
\end{abstract}


\section{INTRODUCTION}\label{sec:introduction}
    Extensive characterization of exoplanets can teach us a lot about their formation process, the composition of their atmosphere or their surface, and their potential habitability. The combination of different observational techniques is essential to sample the widest range of existing exoplanets but also to cross-reference the different obtained data. Among these different observational techniques, long-baseline nulling interferometry offers unique high-angular resolution capabilities that have so-far been out of reach for exoplanets imaging. Even though the technique was first envisioned for the detection of exoplanets in the infrared \citep{bracewell_detecting_1978}, it has mostly been used so far for the detection of debris disks and stellar companions \citep{Hinz1998,Mennesson2011,Mennesson2014,Ertel_2020, martinod_scalable_2021}.
    
    Asgard/NOTT is part of the new visitor instrument suite Asgard for the Very Large Telescope Interferometer (VLTI) observatory in Chile \citep{Martinod2023}. It will be the first nulling interferometer observing in the southern hemisphere and the first one to aim at imaging exoplanets \citep{defrere_hi-5_2018,defrere2022}. Nulling interferometry will achieve high-contrast imaging ($\sim10^{-5}$ stellar attenuation after post-processing) inside the coronagraphic regime \citep{Laugier2023}. This will enable the detection of young giant exoplanets orbiting near the snow line of nearby main-sequence stars ($<$150\,pc distance) [Dandumont et al. in prep.]. The instrument will also perform spectroscopy in the L'-band (3.5 to 4.0\,µm) to help characterizing the atmosphere of the targets. To achieve such contrast in the L'-band, the instrument performs beam combination using a photonic chip made of Gallium Lanthanum Sulfide glass implementing a double Bracewell architecture \citep{gretzinger_towards_2019,Sanny2022}.
    
    The warm optical design of the instrument has been finalized \citep[see][]{Garreau2024} and was mostly used to design the test bench described in this work. The first assembling of the test bench at KU Leuven is then a benchmark in preparation for the first Asgard assembly at Observatoire de la Côte d'Azur (OCA). Section\,\ref{sec:Optomechanical design description} describes the main elements of the test bench and its similarities with the Asgard/NOTT warm opto-mechanical design. Section\,\ref{sec:assembling & alignment strategy} explains the assembling and alignment strategy that is being used. Section\,\ref{sec:first results} gives the current status of the test bench and the first experimental results that have been obtained.

\section{OPTO-MECHANICAL DESIGN OF THE TEST BENCH}\label{sec:Optomechanical design description}
\begin{figure}
        \centering
        \includegraphics[width=\linewidth]{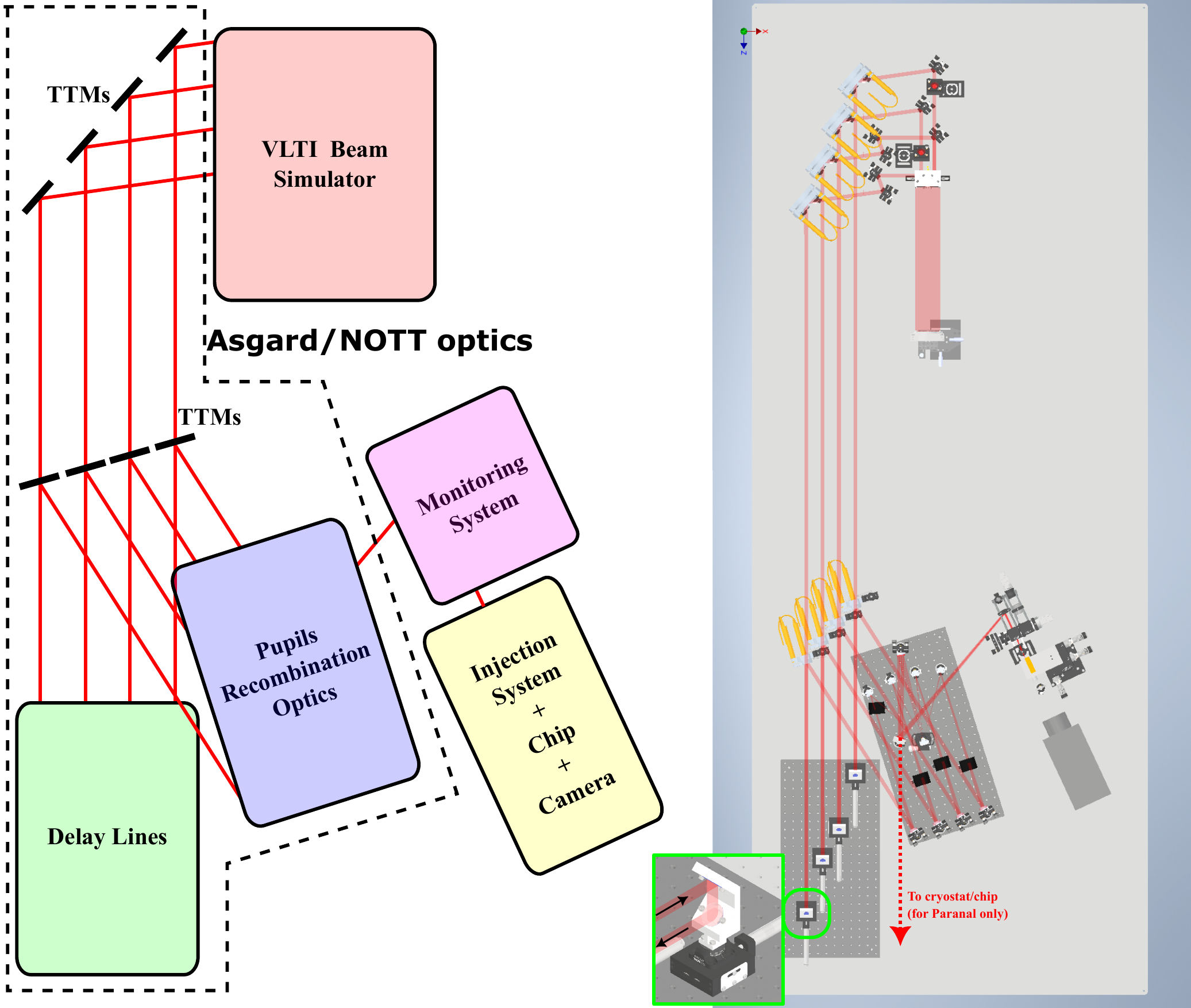}
        \caption{(left) Schematic layout of the Asgard/NOTT test bench at KU Leuven.
        The Asgard/NOTT warm optics (dashed lines) represents the part of the test bench that will be used identically during the Asgard assembling at OCA and Paranal. The main difference is the use of two additional mirrors in the test bench to fold the assembly and fit it on the KU Leuven optical table.
        The injection system is the telecentric lens used with an aperture stop to inject the light into the entrances of the chip. After propagating and interfering inside the chip, the beams are emitted by the different outputs and re-imaged on the infrared camera. The monitoring system is located just before the injection system. Combined with the Tip/Tilt Mirrors (TTMs), it helps with the alignment of the four beams both in the image and pupil planes.
        (right) Top view of the opto-mechanical design of the test bench. The VLTI beam simulator reproduces the four beams as expected from HEIMDALLR and Solarstein. Circled in green is a 3D view of a delay line. The beams are reflected back 40\,mm lower toward the second TTMs.
        The red dotted arrow shows the expected path of the combined beams on the Paranal optical table (without the two additional mirrors).}
        \label{fig:NOTT_test_bench}
    \end{figure}

    A layout of the full opto-mechanical design is shown in Fig.\,\ref{fig:NOTT_test_bench}. It consists of different sections:
    \begin{itemize}
        \item The VLTI beam simulator
        \item The Asgard/NOTT optics
        \item The monitoring system
        \item The injection system, chip, and camera
    \end{itemize}

    \subsection{The VLTI Beam Simulator}\label{sec:VLTI sim}
\begin{figure}
        \centering
        \includegraphics[height=5cm]{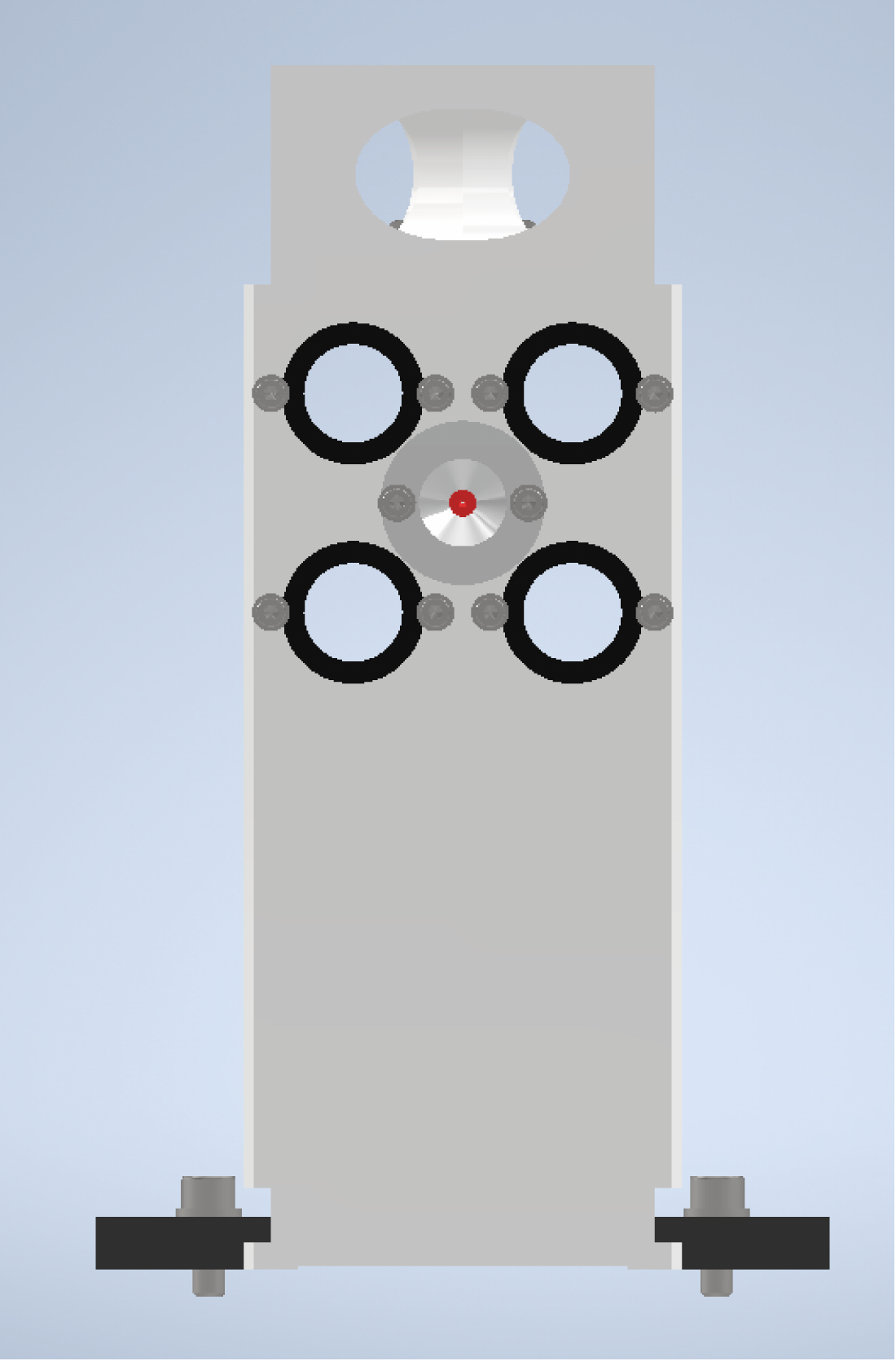}
        \includegraphics[height=5cm]{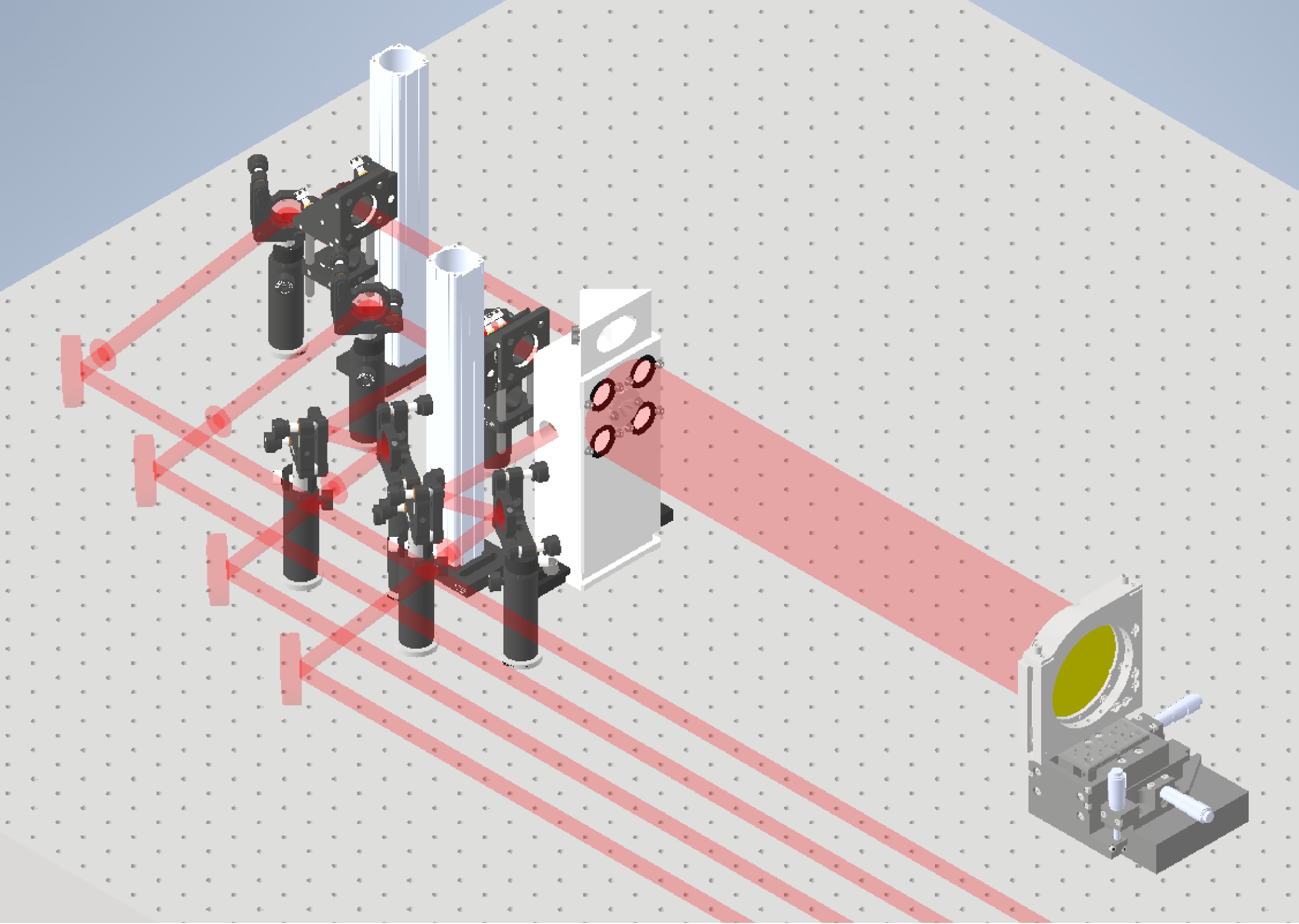}
        \caption{(left) Opto-mechanical design of the Aperture Mask Block with the four 12\,mm-sized apertures and the output of the single-mode fiber at the center (in red). (right) Side view of the opto-mechanical design of the VLTI Beam Simulator.}
        \label{fig:VLTI_sim}
    \end{figure}

        In the Asgard instrument suite, the four beams of the VLTI will first go through the HEIMDALLR module \citep{ireland_image-plane_2018,Taras2024} to perform fringe tracking and wavefront correction with adaptive optics. Four dichroics made of calcium fluoride (CaF$_2$) will then transmit the L'-band toward the NOTT module. The internal calibration and laser alignment will be performed thanks to the Solarstein module, serving as a telescope simulator for the other modules \citep{Taras2024}. \\
        At KU Leuven, we designed and manufactured a VLTI beam simulator to replace the telescope simulator of Asgard (see Fig.\,\ref{fig:VLTI_sim}). It can be used with a green laser or a blackbody source, both via a single-mode fiber generating a gaussian beam. A filter can be added to the blackbody source to transmit only the L'-band. The fiber is connected to the Aperture Mask Block to direct the beam toward a parabola.
        The distance between the parabola and the simulator is chosen so that the four far-field wavefront outputs of the mask are assimilated to four cleaned single-mode gaussian fields of 12\,mm diameter. Mirrors are then used to direct the four beams at the same height and angle as in the Asgard instrument.
    
    \subsection{The Asgard/NOTT Warm Optics}\label{sec:Main optics}
\begin{figure}
        \centering
        \includegraphics[height = 4cm]{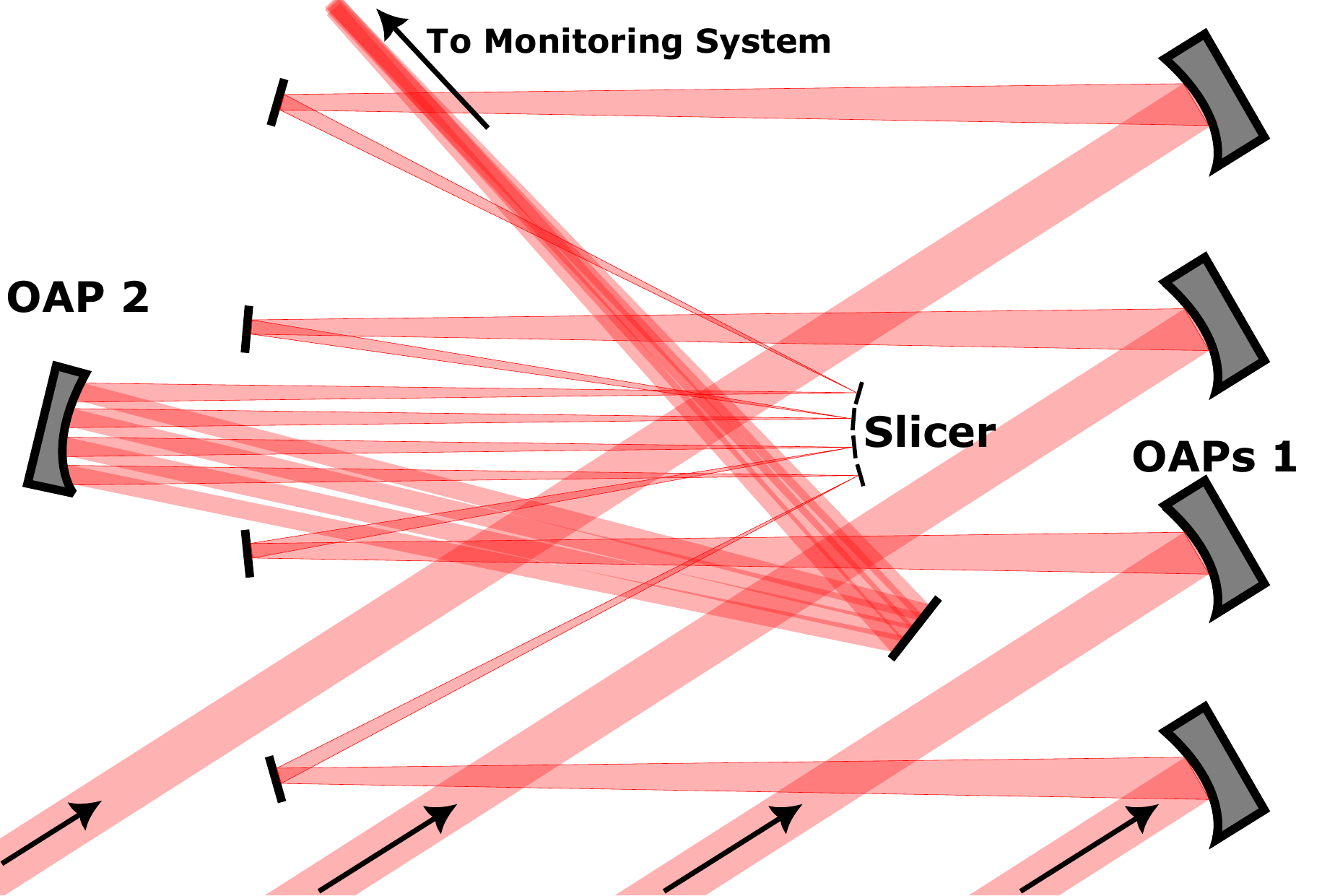}
        \includegraphics[height = 4cm]{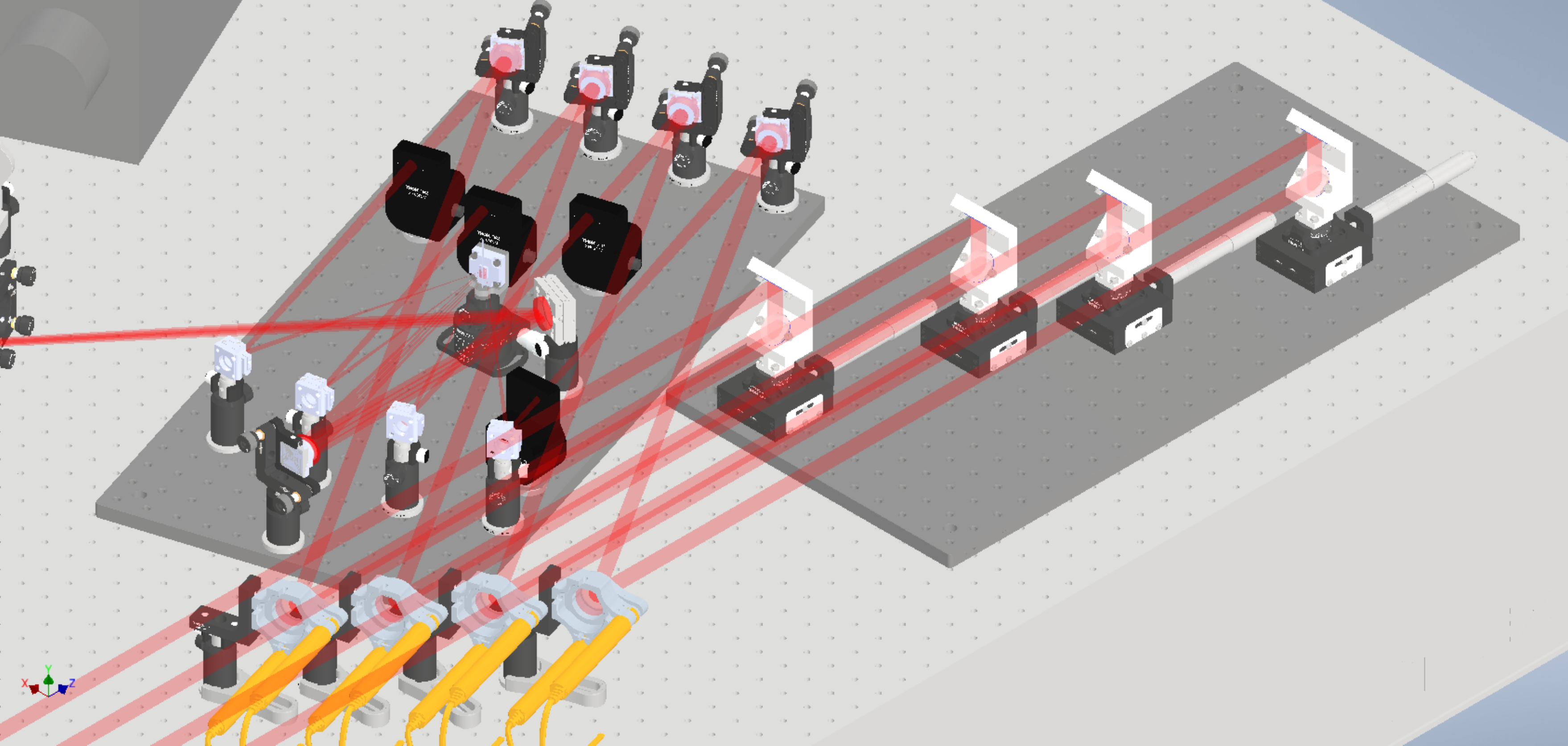}
        \caption{(left) Schematic layout of the Pupils Recombination Optics: several Off-Axis Parabolas (OAPs) focus and re-collimate the beams while four mirror slices, called ``slicer", recombine their pupils at the Injection System. (right) Side view of the opto-mechanical design of the delay lines, the Tip/Tilt Mirrors, and the Pupils Recombination Optics.}
        \label{fig:TB_Recombination_Optics}
    \end{figure}

\begin{figure}
        \centering
        \includegraphics[width=0.3\linewidth]{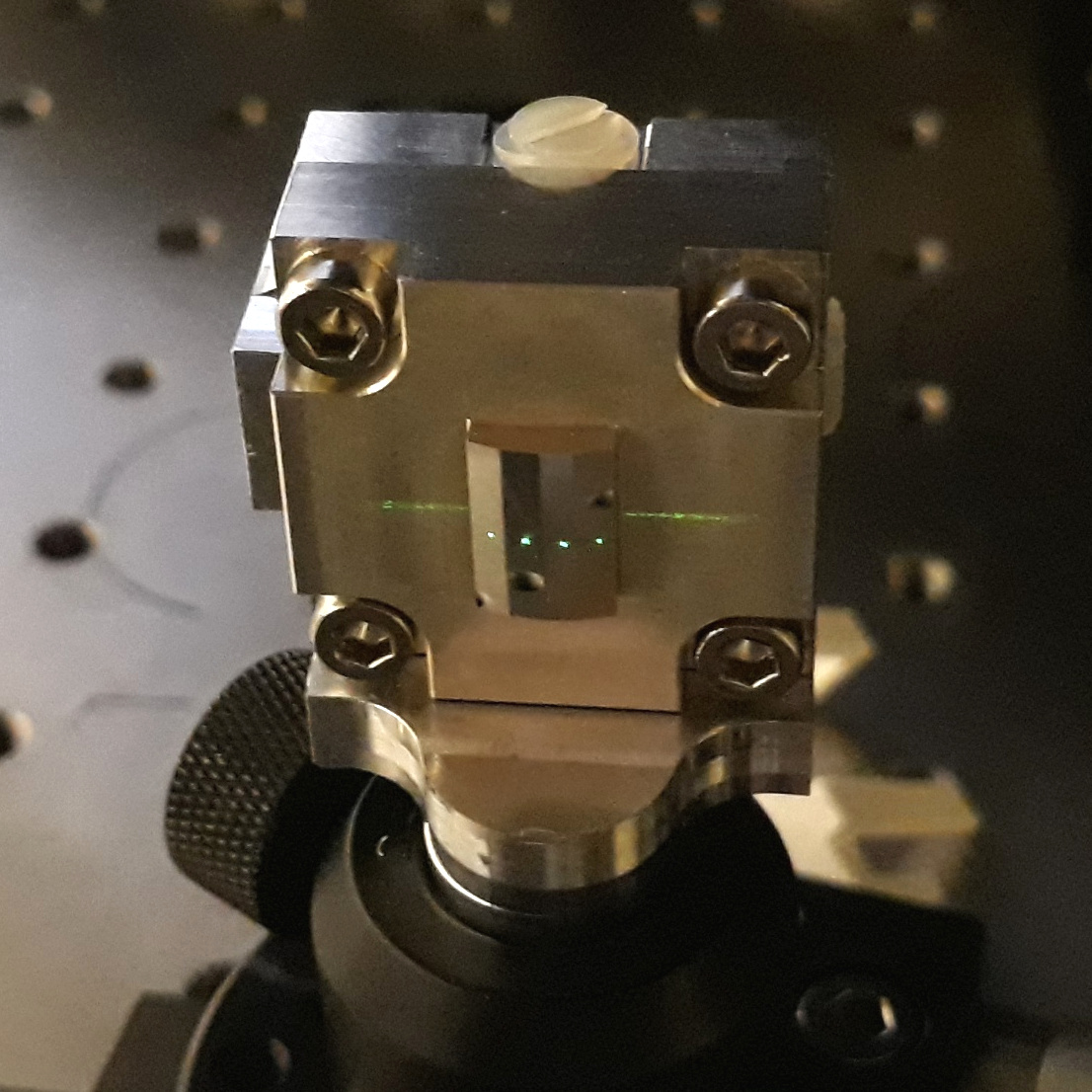}
        \includegraphics[width=0.3\linewidth]{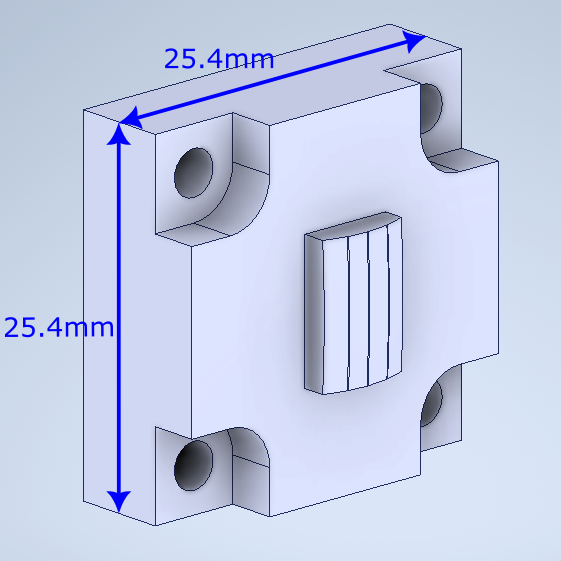}
        \caption{(left) Image of the slicer fixed on its mount. The spots of the beams are visible on the slices with the green laser source. (right) 3D view of the opto-mechanical design of the slicer given by the manufacturer.}
        \label{fig:slicer}
    \end{figure}

        The ``Asgard/NOTT optics" section of the test bench is similar to the one envisioned for the first assembling of the Asgard instrument at OCA before the final installation at Paranal. It corresponds to the design described in \cite{Garreau2024} with two additional folding mirrors to fit into the available KU Leuven optical table. This section is composed of two Tip/Tilt Mirrors (TTMs) and one delay line for each beam. Using four mirror slices, called ``slicer", and Off-Axis Parabolas (OAPs), the system recombines the four pupils of the beams before the injection in the chip (see Fig.\,\ref{fig:TB_Recombination_Optics}-\ref{fig:slicer}). The combined pupil plane is located at the same position as the aperture stop before the injection lens (see Fig.\,\ref{fig:Monitoring+injection}). The delay lines and Pupils Recombination Optics are both assembled on their own respective breadboards to facilitate their moving to OCA and later Paranal. \\
        The two additional mirrors used for the test bench are folding the combined beam after OAP$_2$ but they don't impact its pupil and image planes. The assembly and alignment of this sub-system stays then equivalent and therefore relevant for the Asgard assembly. Table\,\ref{tab:NOTT warm optics list} gives the list of Asgard/NOTT warm optics used in the test bench and their surface quality. The dichroics have been ordered and will be implemented in the test bench in the near future.

    \begin{table}
        \centering
        \caption{List of the Asgard/NOTT warm optics used in the test bench at KU Leuven. The surface quality is given in peak-to-valley (P-V) or root mean square (RMS). The flat mirrors include the optics of the TTMs, the delay lines, and the folding mirrors.}
        \begin{tabular}{c c c}
            \hline
            Name of the optic & Manufacturer & Surface quality (@633\,nm) \\ \hline
            Dichroics & Rocky Mountain Instrument Co. & $\lambda$/4 P-V \\
            Flat mirror & Edmund Optics & $\lambda$/20 P-V \\
            Off-Axis Parabolas & Wielandts UPMT & $\lambda$/40 RMS \\
            Slicer & Wielandts UPMT & $\lambda$/40 RMS \\\hline
        \end{tabular}
        \label{tab:NOTT warm optics list}
    \end{table}

          \begin{table}
        \centering
        \caption{Specifications for the visible cameras used in the monitoring system.}
        \begin{tabular}{c c}
            \hline
            Model number & PHX064S-MC \\
            Power Supply & Power over Ethernet \\
            Type of Sensor & Progressive Scan CMOS \\
            Pixels (H $\times$ V) & 3072 $\times$ 2048 \\
            Pixel size (µm) & 2.40 $\times$ 2.40 \\
            Exposure time & 30\,µs - 10\,s \\
            Imaging Sensor & Sony IMX178 \\ \hline
        \end{tabular}
        \label{tab:visible cameras}
    \end{table}
    
    \subsection{The Monitoring System}\label{sec:Monitoring system}
        To help with the alignment of the chip and the control of long term stability, a cage system is installed before the chip. This sub-system is meant to work with visible light generated by the blackbody source on the test bench or the Solarstein module in Asgard. Two visible cameras are used to monitor both the image and pupil planes of the four beams simultaneously. The specifications of the cameras are listed in Table\,\ref{tab:visible cameras}. The feedback from the two cameras is used in coordination with the TTMs to help aligning the pupil planes with the aperture stop and the spots at the entrances of the chip. The visible light can be directed to the monitoring system thanks to a vertical translation stage below the first beamsplitter. When in the upper position, the light is fed to the visible cameras while in the lower position, the light is not impacted by the beamsplitter.

    \subsection{The Injection and Imaging of the Outputs}\label{sec:Injection and Imaging outputs}
\begin{figure}
        \centering
        \includegraphics[width = 0.4\linewidth]{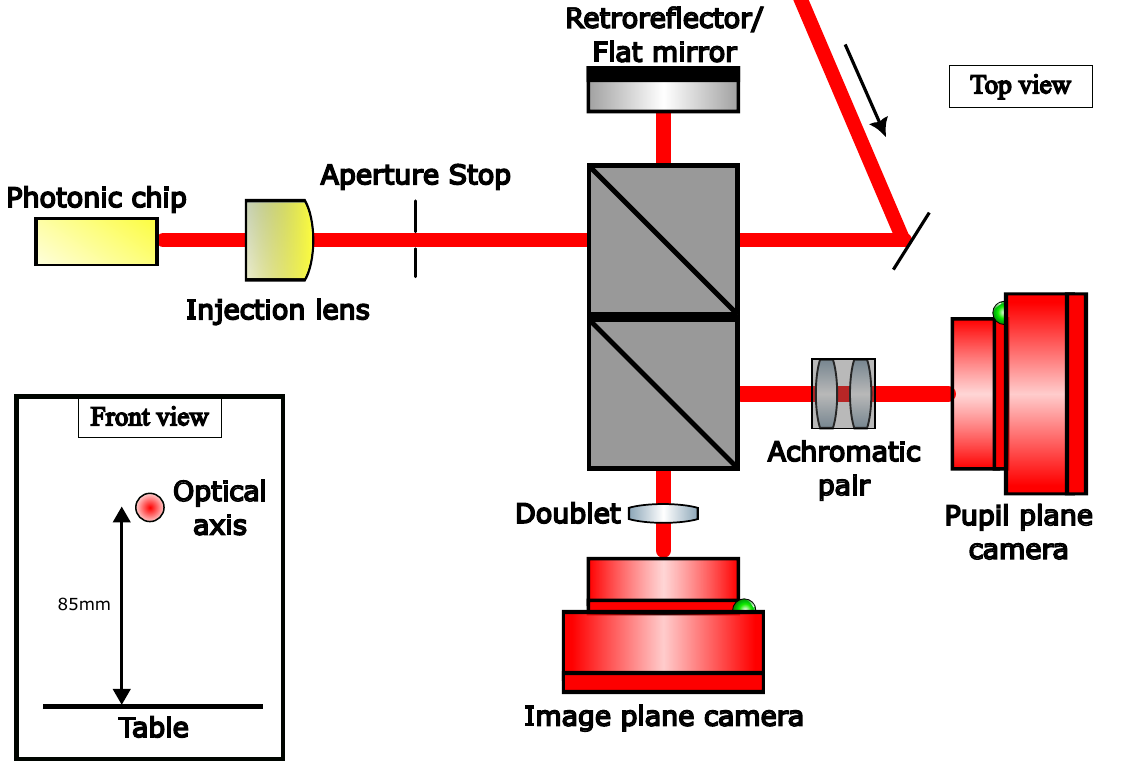}
        \includegraphics[width=0.55\linewidth]{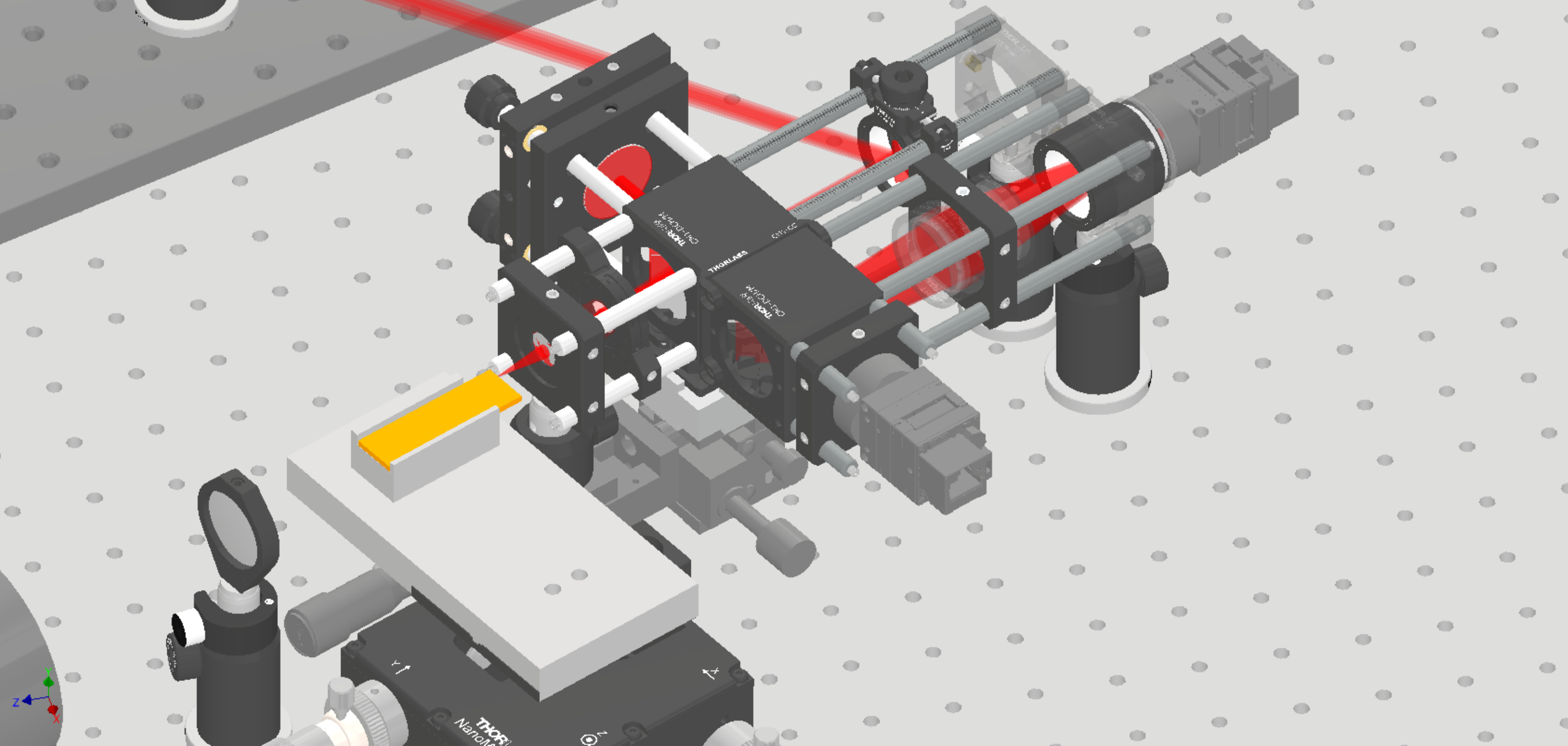}
        \caption{(left) Schematic layout of the monitoring system and injection system. (right) Side view of the opto-mechanical design of the monitoring system and injection system. The collimation lens and infrared camera can be seen at the left of the chip. A vertical translation stage is located below the first beamsplitter to move it up or down. This enables to decide when to feed the visible light to the cameras.}
        \label{fig:Monitoring+injection}
    \end{figure}

        A photonic chip similar to the one that will be used in Asgard/NOTT is used in this test bench. This chip is a prototype manufactured at Macquary University and previously tested at Universität zu Köln.
        The injection of the four beams in the chip is performed with a telecentric aspherical lens made of zinc selenide (ZnSe) (see Fig.\ref{fig:Monitoring+injection}). Figure\,\ref{fig:chip} shows a schematic layout of a nuller block inside the photonic chip, dedicated to the beam combination and interferometry. The four entrances, one for each beam, are separated by 125\,µm. An S-bend of 1.8\,mm amplitude allows to avoid stray light contamination. Two stages of directional couplers allow then to interfere the four beams. The eight different outputs are emitted at the end of the chip, it includes: four photometric outputs emitting $\sim20\,\%$ of the injected light, two constructive outputs emitting mainly the on-axis stellar light, and two destructive outputs emitting the potential off-axis light from exoplanets or dust. These eight outputs are re-imaged on an infrared camera InfraTec ImageIR® 8310.
        
\section{ASSEMBLING AND ALIGNMENT STRATEGY}\label{sec:assembling & alignment strategy}
    This section describes the strategy tested during the first assembling and alignment of Asgard/NOTT at KU Leuven. This strategy is meant to be a benchmark for the assembling of NOTT within the Asgard instrument at OCA, and later at Paranal. The different steps of this assembling are:
    \begin{itemize}
        \item \textbf{Positioning of the optical mounts} with $\sim$1\,mm precision
        \item \textbf{First alignment} using a bright visible light source
        \item \textbf{Second alignment and injection} in the infrared using a blackbody source
    \end{itemize}
    
    \subsection{Positioning of the Optical Mounts}
\begin{figure}
            \centering
            \includegraphics[width=0.7\linewidth]{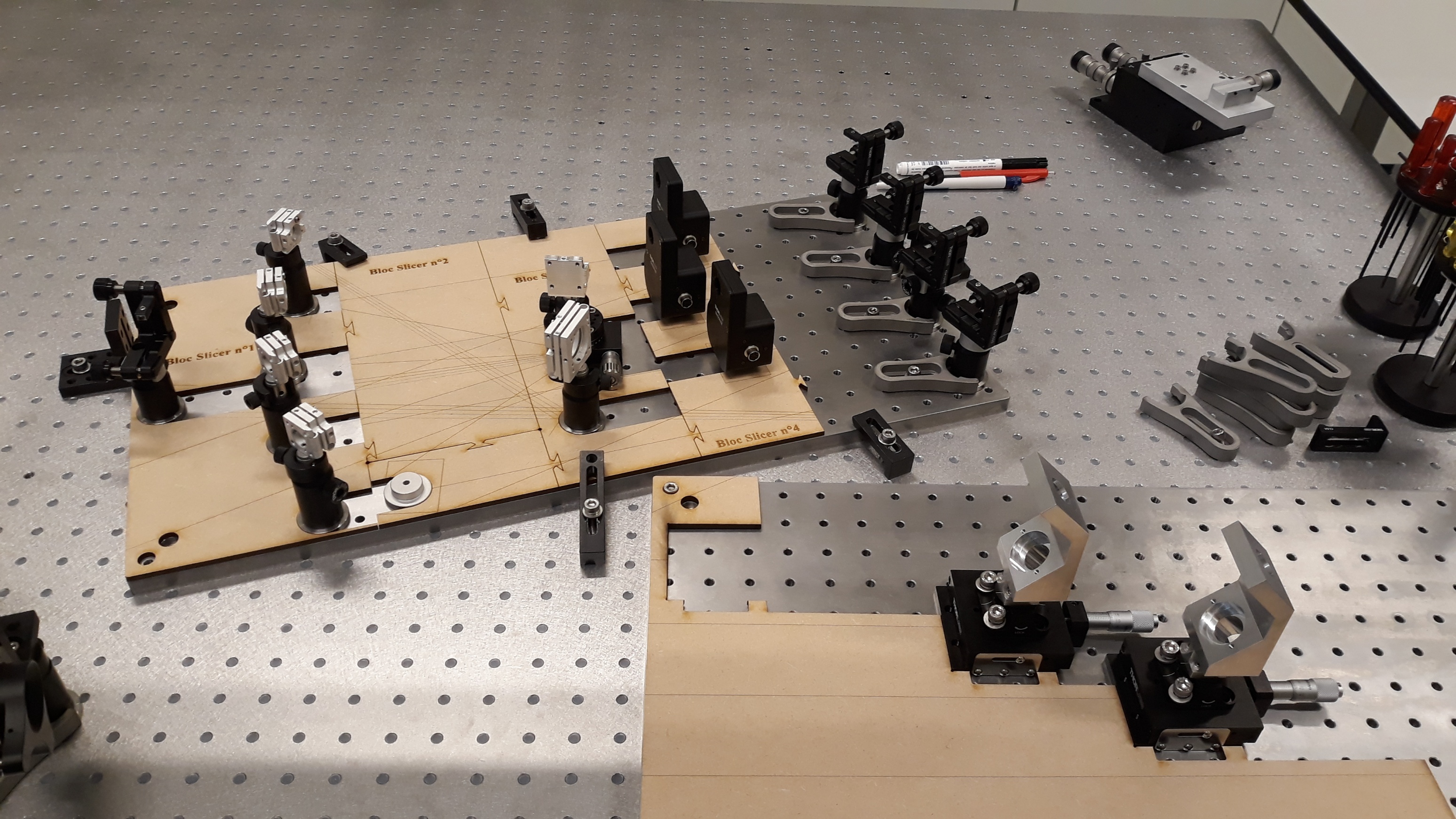}
            \caption{Example of patterns made from MDF with laser cutting. The two patterns shown here are used to position the Pupils Recombination Optics and the delay lines.}
            \label{fig:alignment bloc}
        \end{figure}

        The objective of this first step is to obtain a precision of $\sim$1\,mm on the position of the optical mounts. To achieve this level of precision, a laser cutting machine realizes patterns on a Medium-Density Fiberboard (MDF) that are then used for positioning the mounts. The correct patterns to make are obtained from the opto-mechanical design of the test bench described in Sec.\,\ref{sec:Optomechanical design description}. Figure\,\ref{fig:alignment bloc} shows examples of patterns realized for the Pupils Recombination Optics and the delay lines. The precision of the laser cutting is $\sim$50\,µm, which enables to achieve the $\sim$1\,mm precision target.

    \subsection{First Alignment}
\begin{figure}
        \centering
        \includegraphics[width=0.32\linewidth]{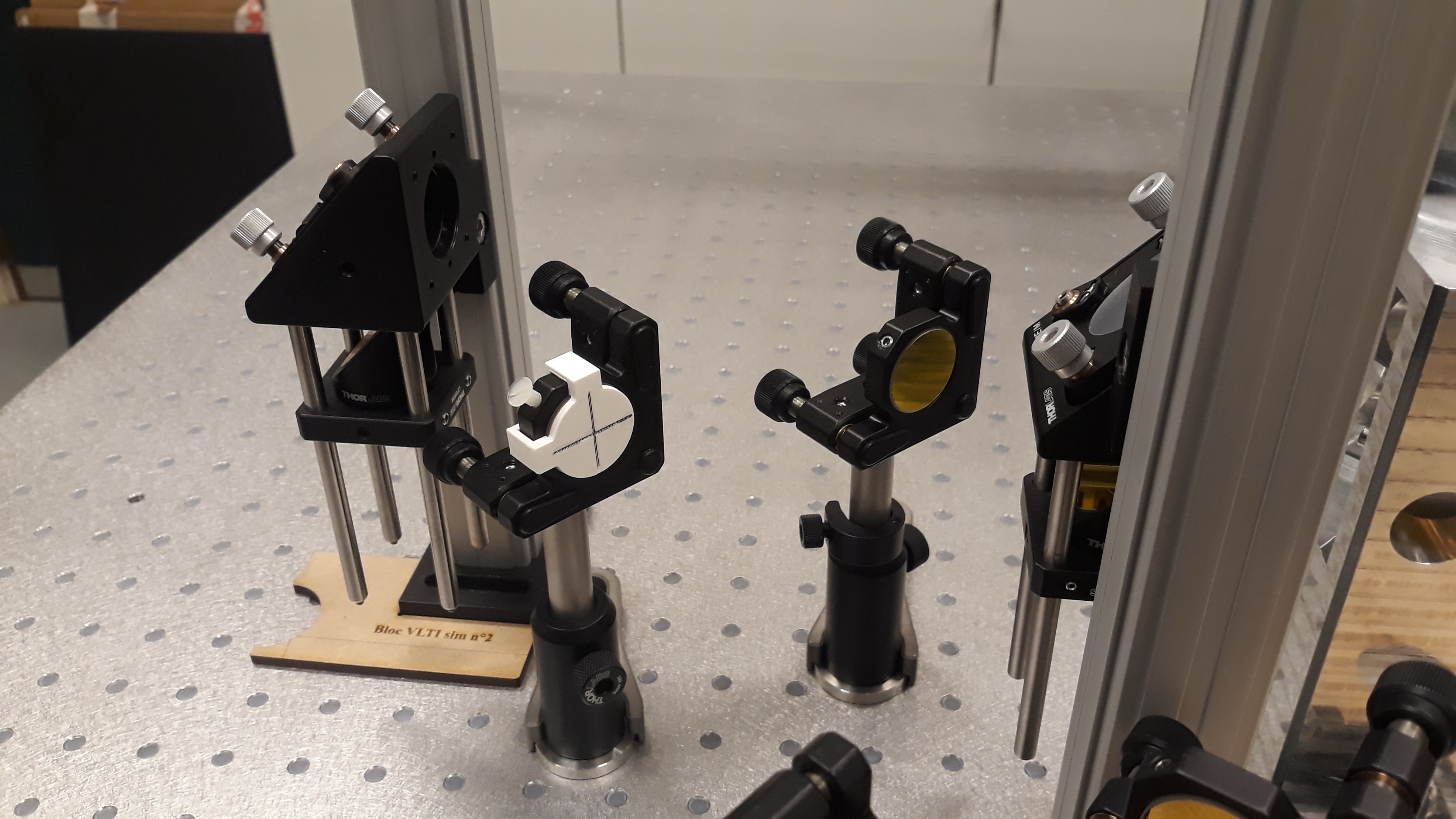}
        \includegraphics[width=0.32\linewidth]{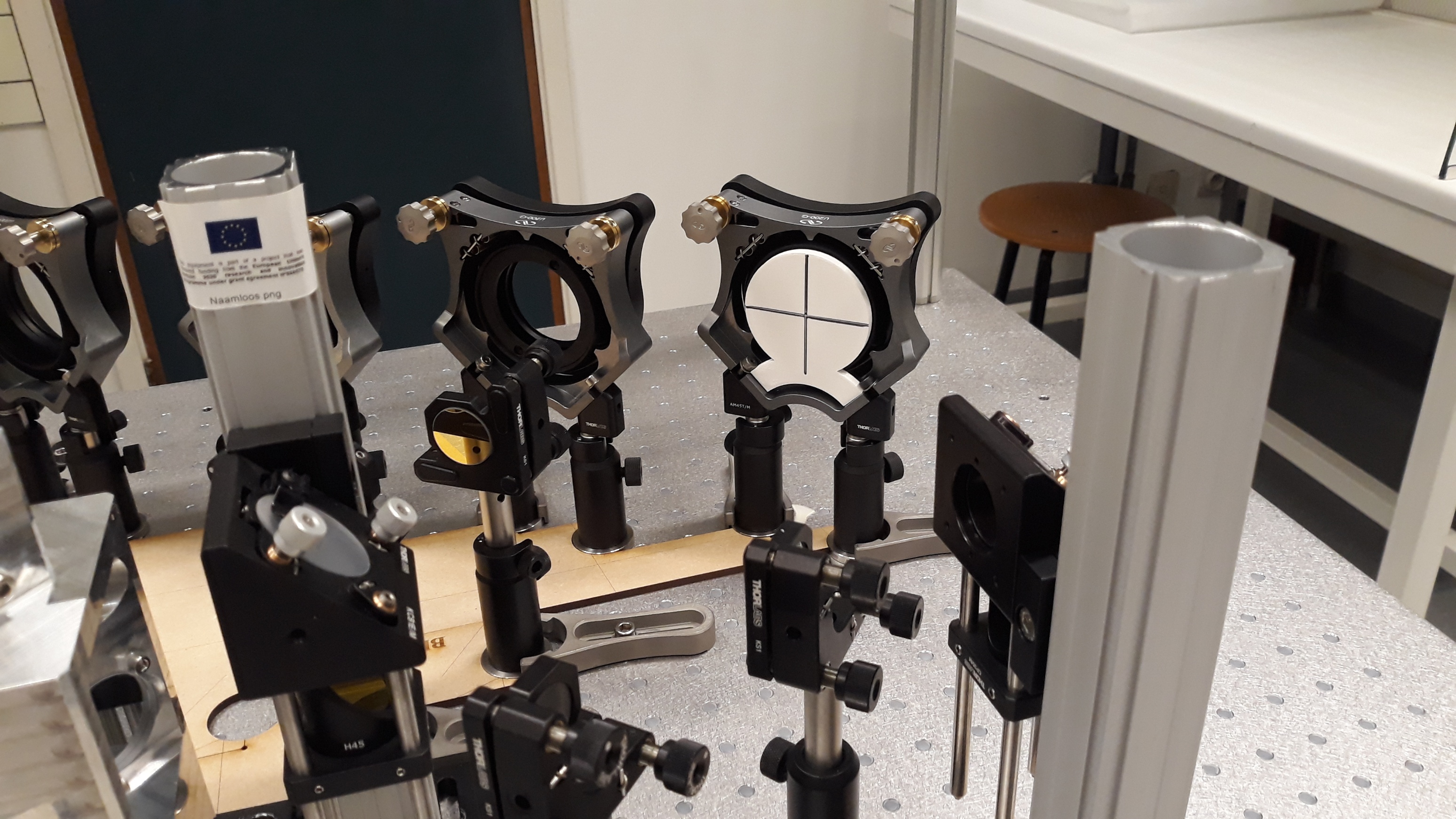}
        \includegraphics[width=0.32\linewidth]{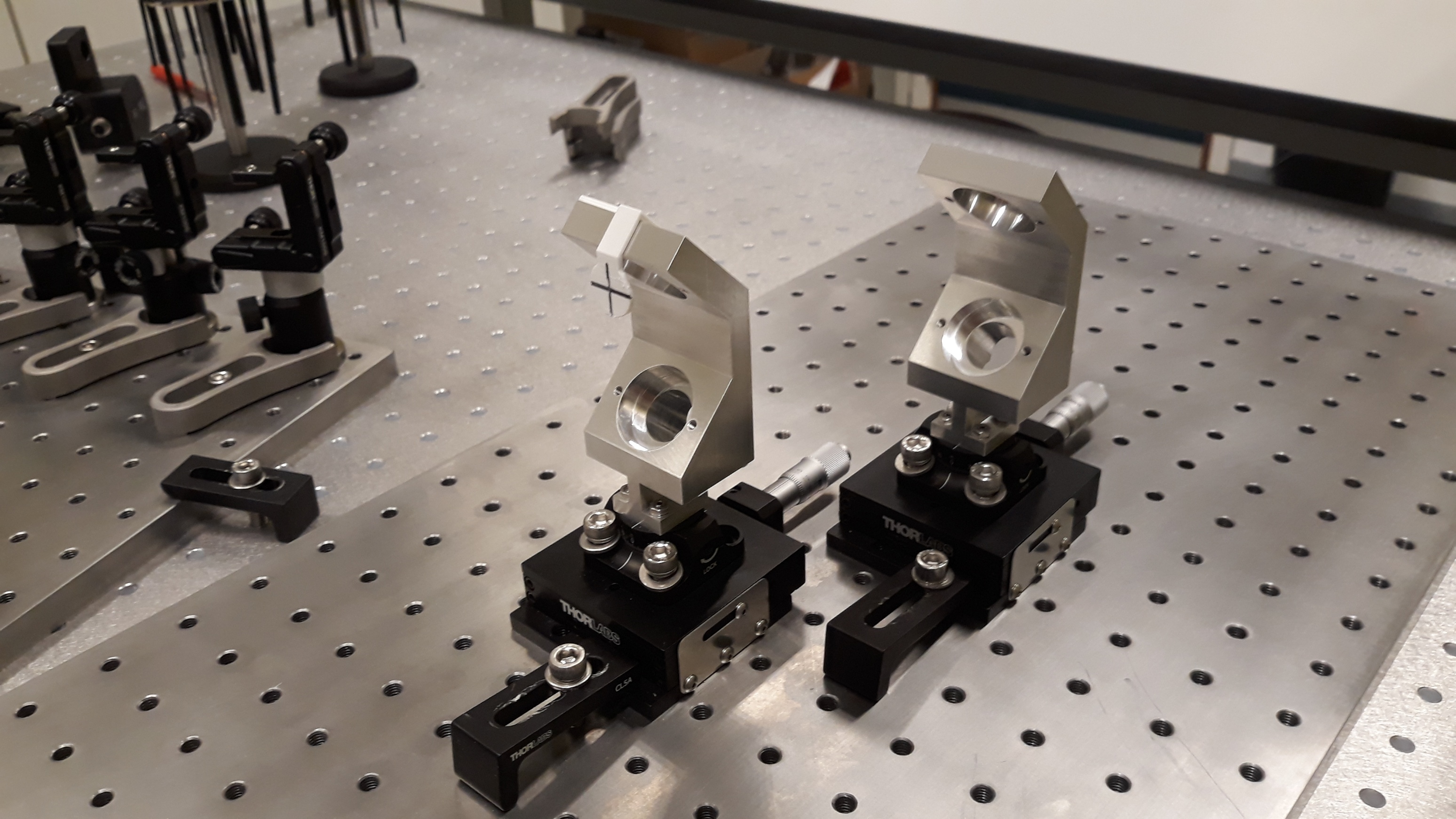}
        \caption{Example of 3D printed targets used for the first alignment in visible light.}
        \label{fig:Target examples}
    \end{figure}

        The first alignment step is performed using a green laser source. A single-mode fiber adapted for visible light is coupled with the Aperture Mask Block (see Fig.\,\ref{fig:VLTI_sim}) and generates a gaussian beam. After the first positioning of the optics, the goal is to direct the light to the center of each optics using the degrees of freedom of the previous mount. To help with the centering of the light on the optics, custom pieces are realized using a 3D printer. These custom pieces are adapted for each type of mounts in the system and have a target on it. Figure\,\ref{fig:Target examples} shows some examples of 3D printed targets.\\

        \begin{figure}
            \centering
            \includegraphics[width=0.75\linewidth]{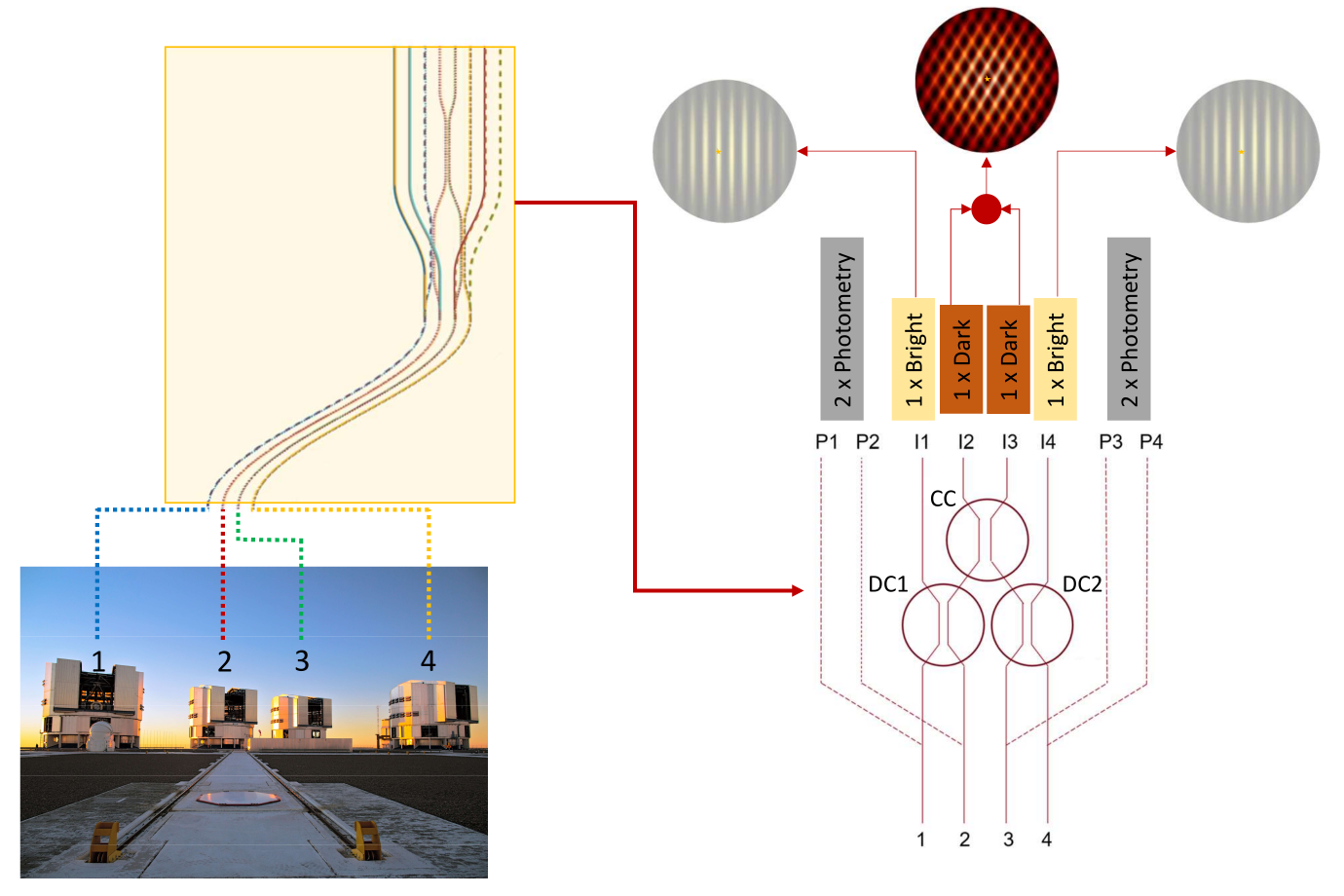}
            \caption{Schematic layout of the photonic chip designed for Asgard/NOTT. On the left, the reproduction of the entire chip with inputs coming from the VLTI telescopes. The S-bend used to reduce stray light after the input is visible. On the right, a closer view of the couplers and the eight outputs. The photometric outputs P1, P2, P3, and P4 emit 20\,\% of the injected light. The bright interferometric outputs I1 and I4 mainly emit the on-axis stellar signal. The dark interferometric outputs I2 and I3 emit the off-axis signal corresponding to possible exoplanets or disks. Source: \cite{Sanny2022}.}
            \label{fig:chip}
        \end{figure}
        
\begin{figure}
        \centering
        \includegraphics[height=4cm]{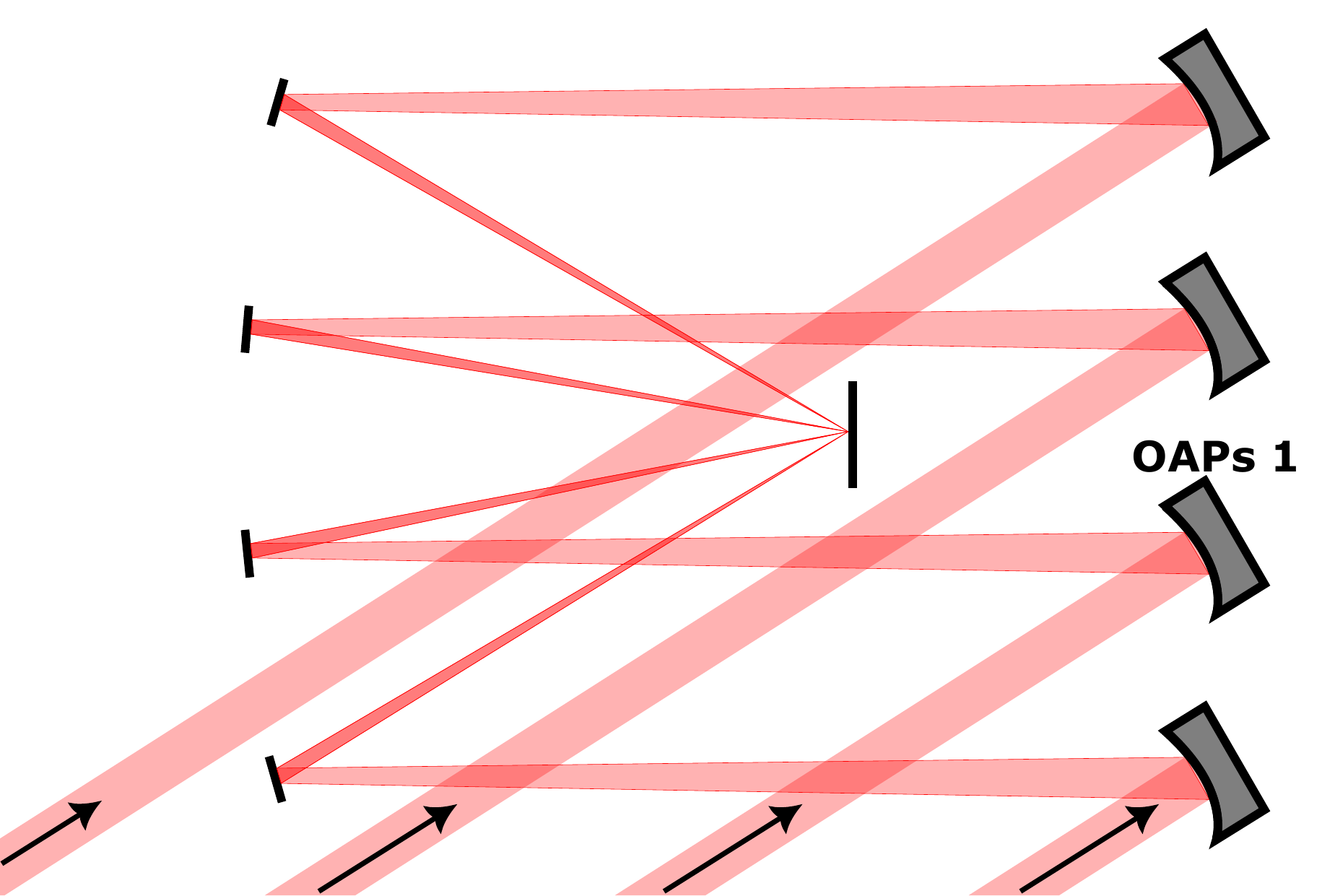}
        \caption{Schematic layout of the Pupils Recombination Optics during the first alignment phase. The slicer at the center is replaced by a single flat mirror. Thanks to the symmetry of the assembly, the beams from the folding mirrors on the left should be reflected to the opposite folding mirrors. This ensures a correct positioning and orientation of the slicer mount when the slicer is inserted.}
        \label{fig:slicer alignment}
    \end{figure}

        For the alignment of the slicer, a specific procedure is required. The slicer is indeed composed of four mirror slices with a convex shape, which complicates its alignment with the optics upstream. To constrain the angle of the slicer mount, it is first replaced by a flat mirror. Figure\,\ref{fig:slicer alignment} shows a sketch of the assembly. The OAPs$_1$, folding mirrors, and slicer are horizontally symmetric; thus when the slicer is replaced with a flat mirror, the beams should be reflected to the opposite mirrors and follow a reverse optical path. When this is obtained, the flat mirror can be swapped with the slicer on the same mount, ensuring that the constrained angle is kept during the process. The folding mirrors can then be used to direct the beams on their corresponding mirror slices.\\
        
        During this phase, the monitoring system is also adjusted thanks to the green laser. The first beamsplitter is moved upwards to feed the two visible cameras. The adjustment of the monitoring system (see Fig.\,\ref{fig:Monitoring+injection}) follows these steps:
        \begin{itemize}
            \item Setting the doublet+image plane camera at infinity
            \item Using the retro-reflector to mark the positions of the four spots from the beams
            \item Replacing the retro-reflector with a flat mirror
            \item Using the marked position of the spots to align the flat mirror
            \item Setting the achromatic pair to image the aperture stop on the pupil plane camera
        \end{itemize}

\begin{figure}
        \centering
        \includegraphics[width=0.4\linewidth]{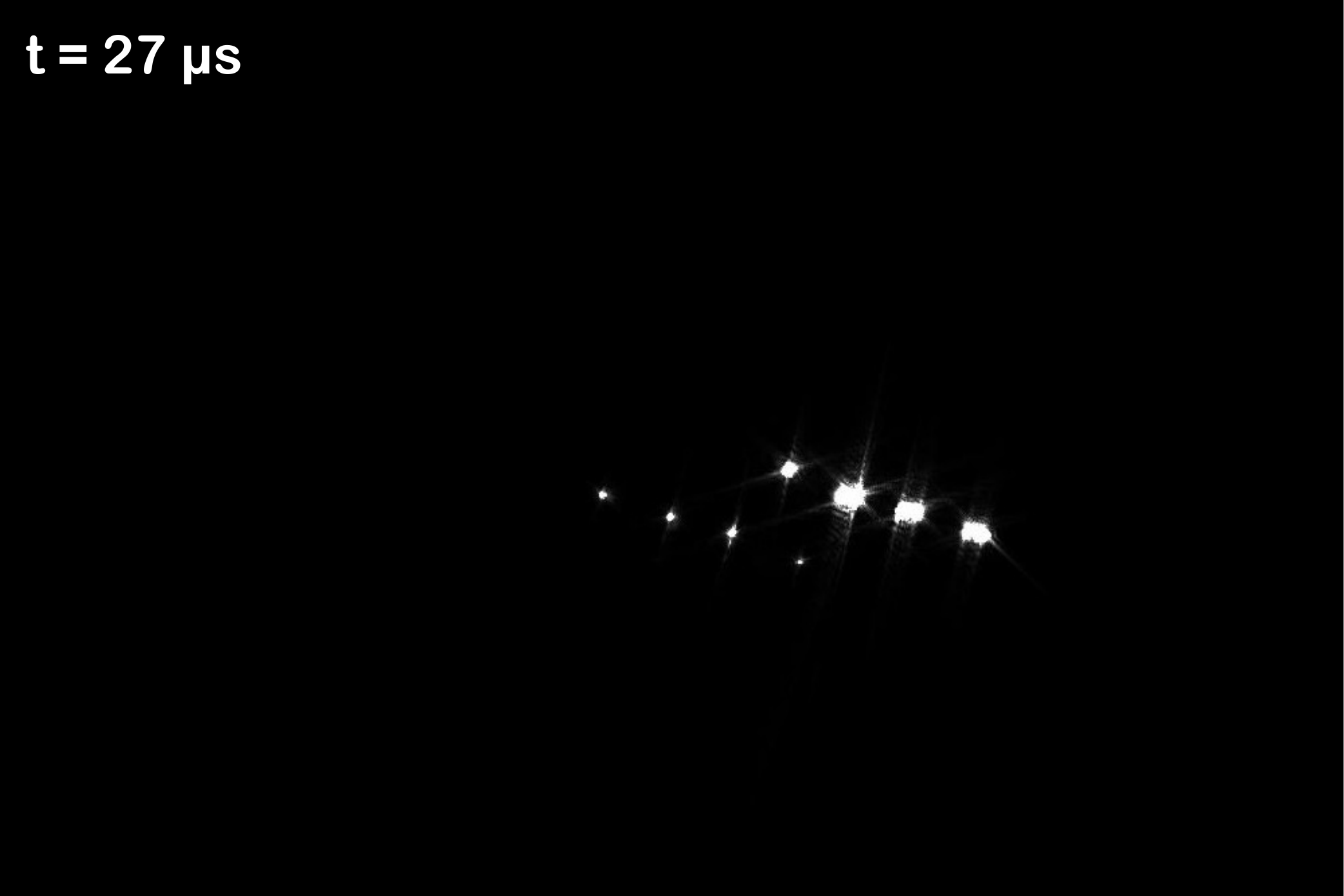}
        \includegraphics[width=0.4\linewidth]{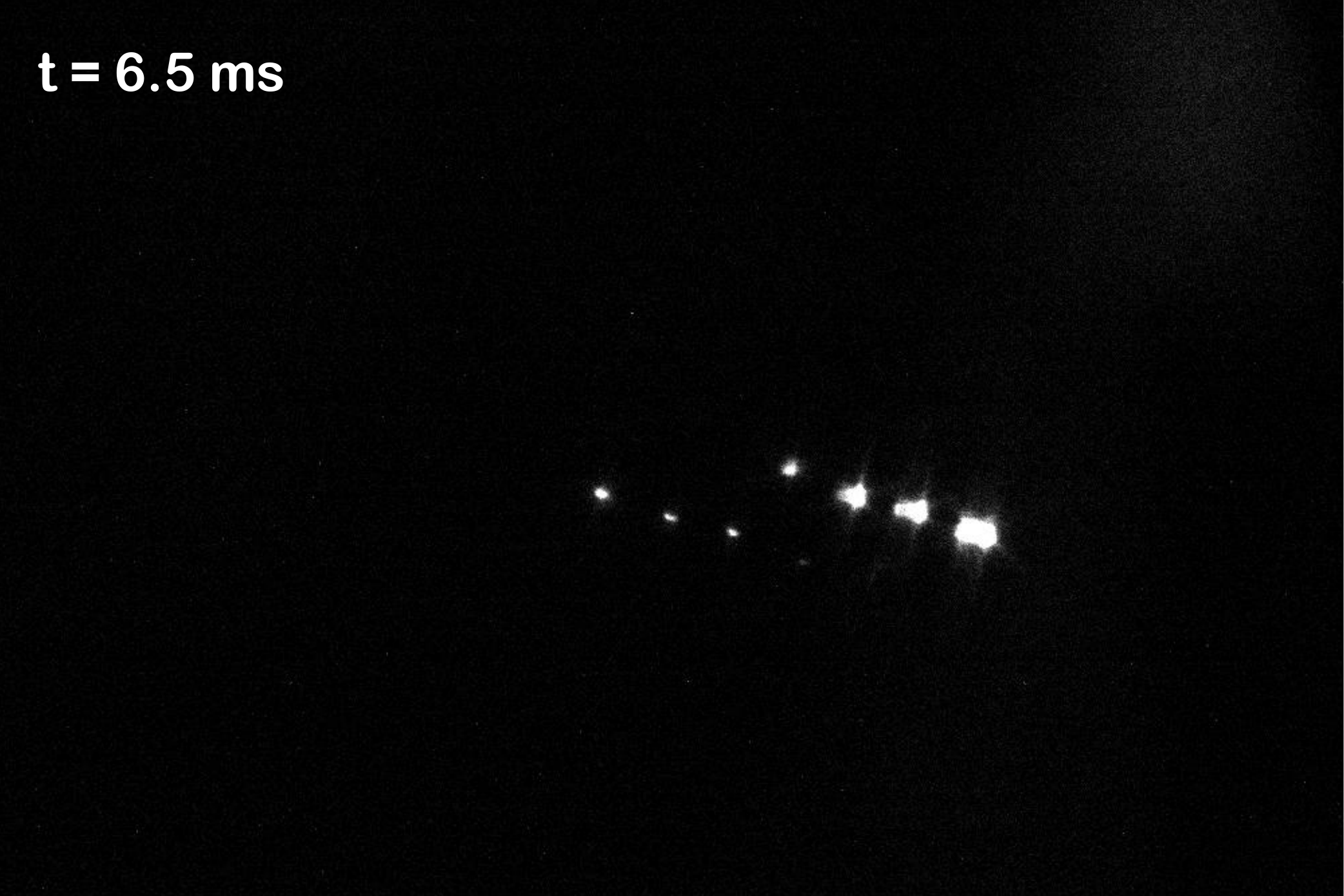}
        \includegraphics[width=0.4\linewidth]{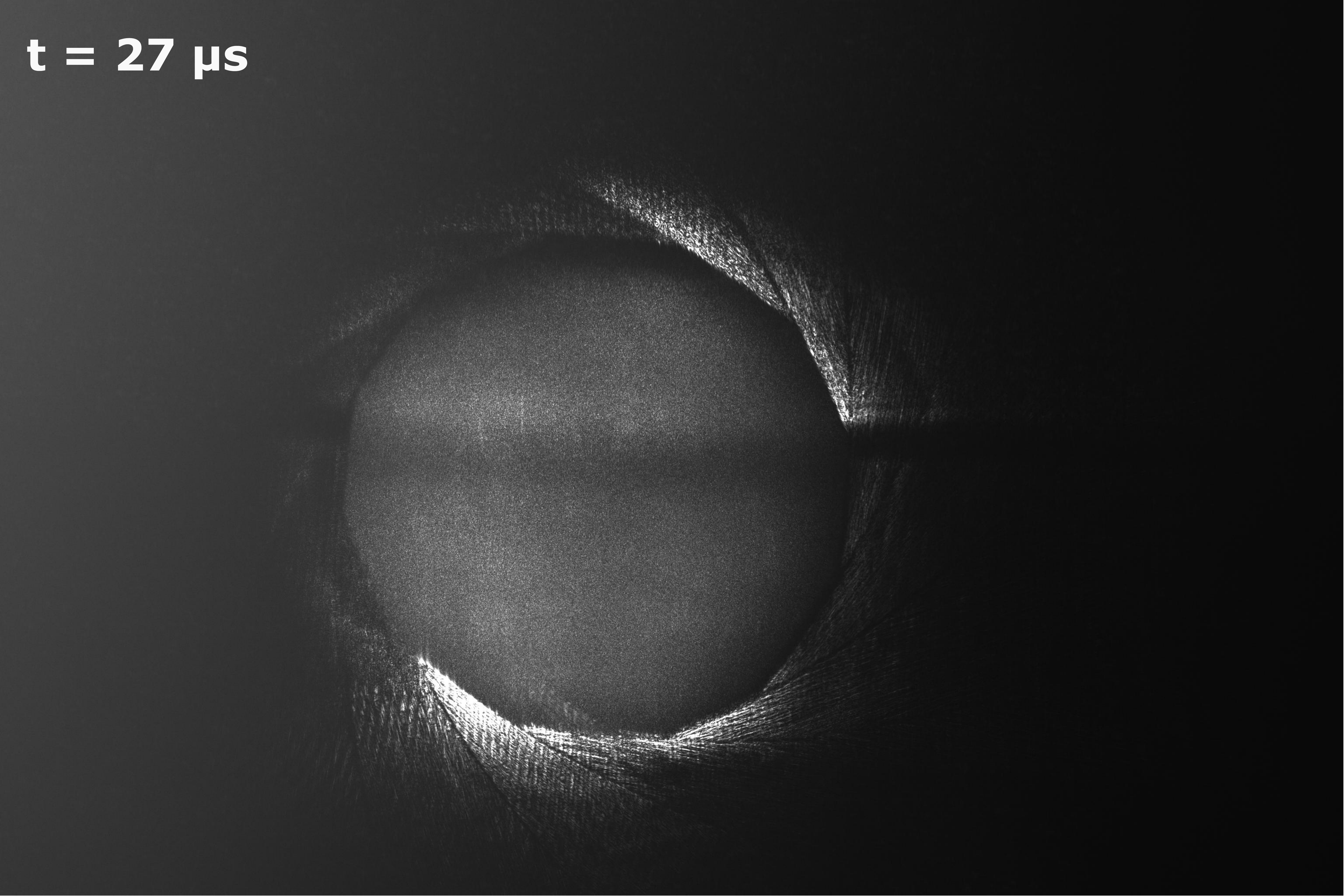}
        \includegraphics[width=0.4\linewidth]{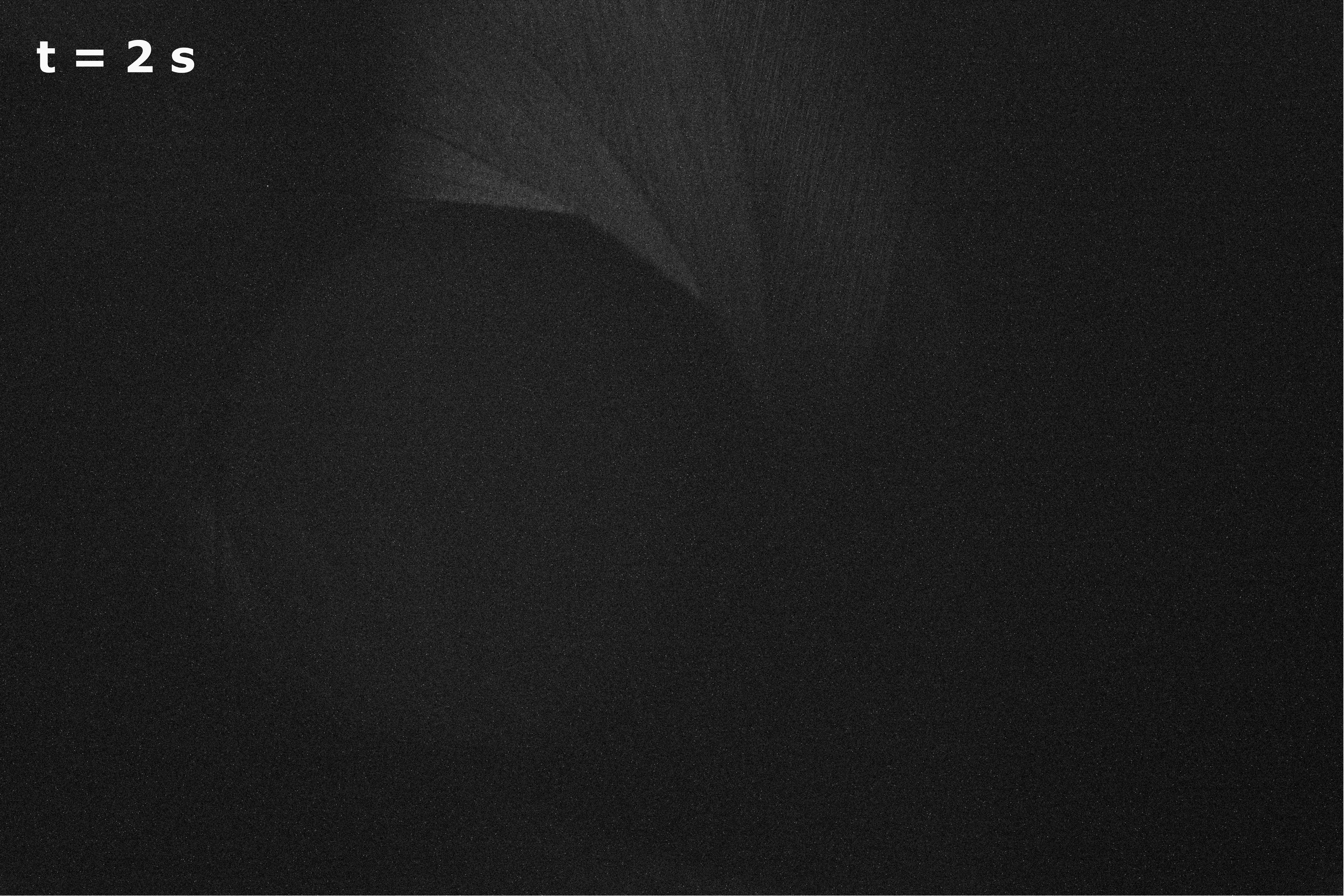}
        \caption{Results of the monitoring cameras during the first and second alignment phase, with a green laser (left) and a blackbody source (right) respectively. (top) Image plane camera: the four spots are observed with their fainter ghost reflections generated by the beamsplitter. (bottom) Pupil plane camera. In the bottom left image, reflections of the beams on the aperture stop can be seen at the outer edge of the aperture. The bottom right image shows that no reflection can be seen when using the blackbody source because of the faintness of the light. A more reflective aperture stop would be needed to observe it. For every images, the integration time is indicated in the top left corner.}
        \label{fig:monitoring cameras}
    \end{figure}

        Figure\,\ref{fig:monitoring cameras} shows the outputs of the two monitoring cameras after the adjustment with the green laser. The reflection of the beams on the surface of the stop can be used to improve their alignment with it.

    \subsection{Second Alignment and Injection}
        The last step corresponds to a second alignment of the test bench using a blackbody source emitting mainly in the infrared. During this phase, the chip support is used to inject the four beams in the entrances of the chip. The TTMs are then used to fine tune the alignment of the beams with the aperture stop and their corresponding entrances at the chip.
        The monitoring system supports this phase by inspecting the image plane and the combined pupil plane. Figure\,\ref{fig:monitoring cameras} shows that the position of the four spots in the image plane is similar when switching from the green laser to the blackbody source. This confirms the consistency of the VLTI beam simulator when performing the switching.
        
        In the image of the pupil plane (see Fig.\,\ref{fig:monitoring cameras}), the green laser is bright enough to observe the reflections on the aperture stop and use it to align the pupils of the beam. When switching to the blackbody, however, the visible light becomes too faint and no reflection can be observed even with a high exposure time (2\,s). A more reflective surface would be needed to possibly use the reflections for pupil alignment.

    \begin{figure}
        \centering
        \includegraphics[width=\linewidth]{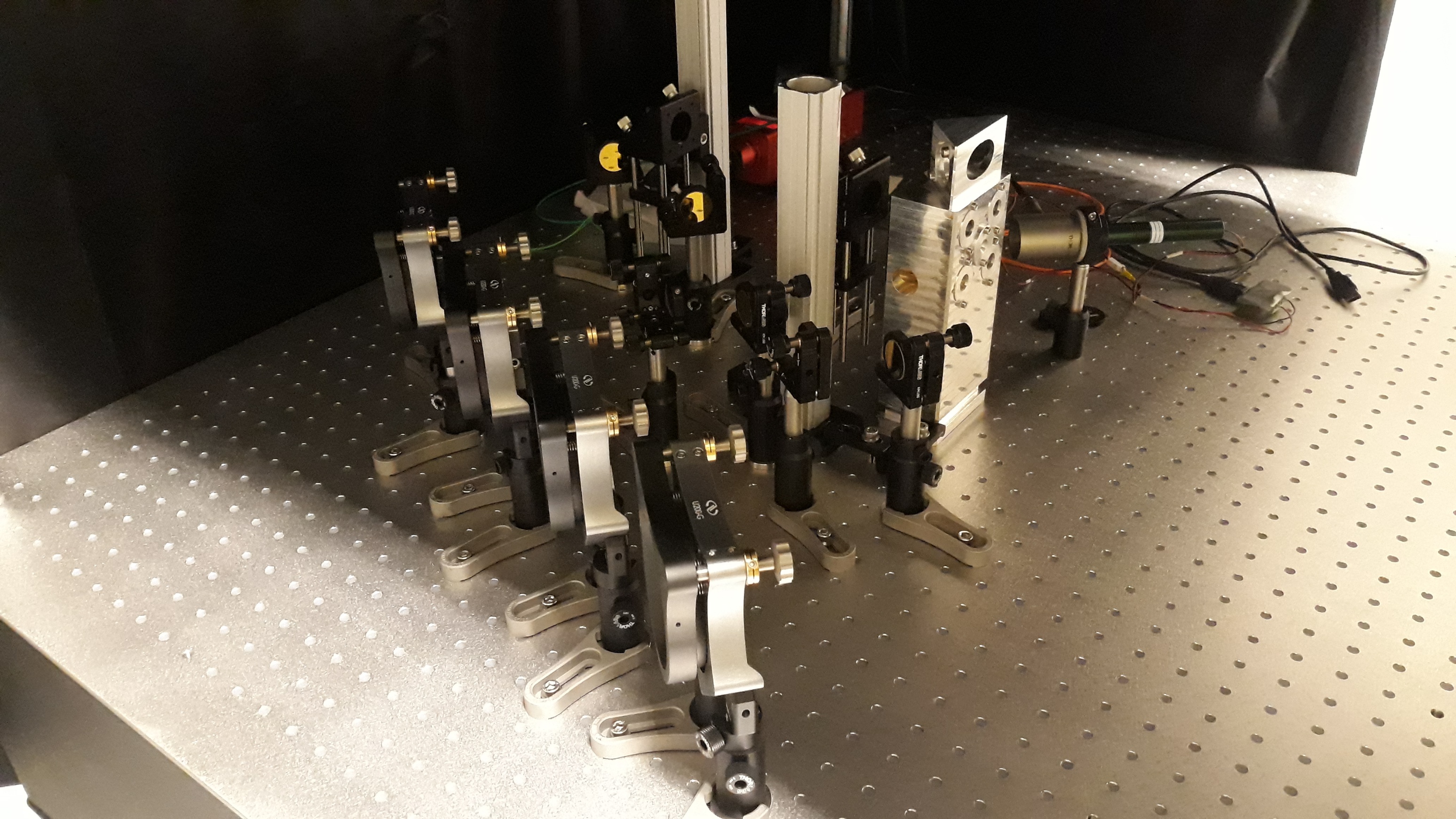}
        \includegraphics[width=\linewidth]{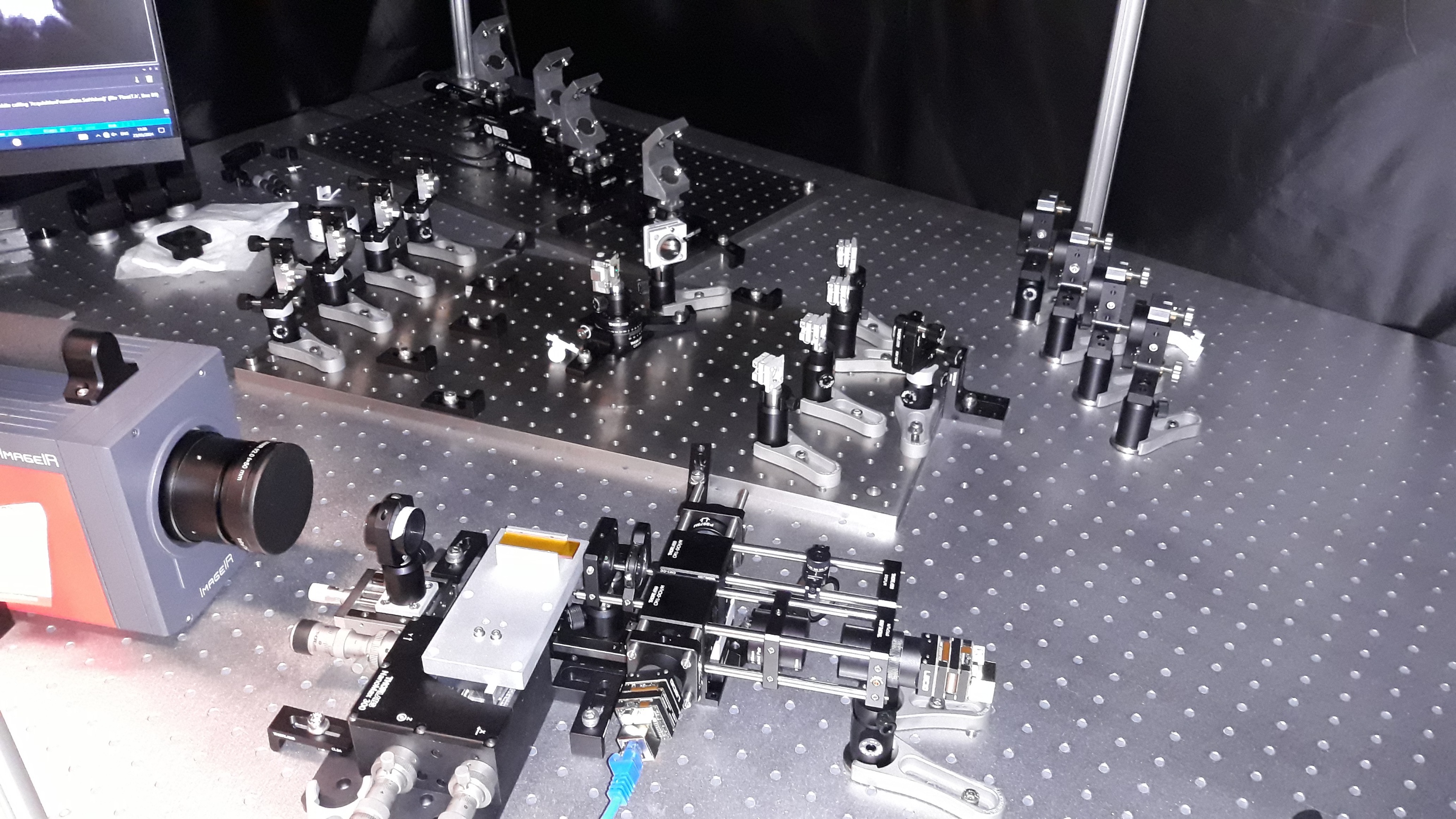}
        \caption{Images of the current test bench assembly at KU Leuven. (top) VLTI beam simulator and first TTMs. (bottom) Overview of the Asgard/NOTT optics, monitoring system, photonic chip, and infrared camera.}
        \label{fig:lab status}
    \end{figure}

\section{STATUS AND FIRST EXPERIMENTAL RESULTS}\label{sec:first results}
\begin{figure}
        \centering
        \includegraphics[width=0.9\linewidth]{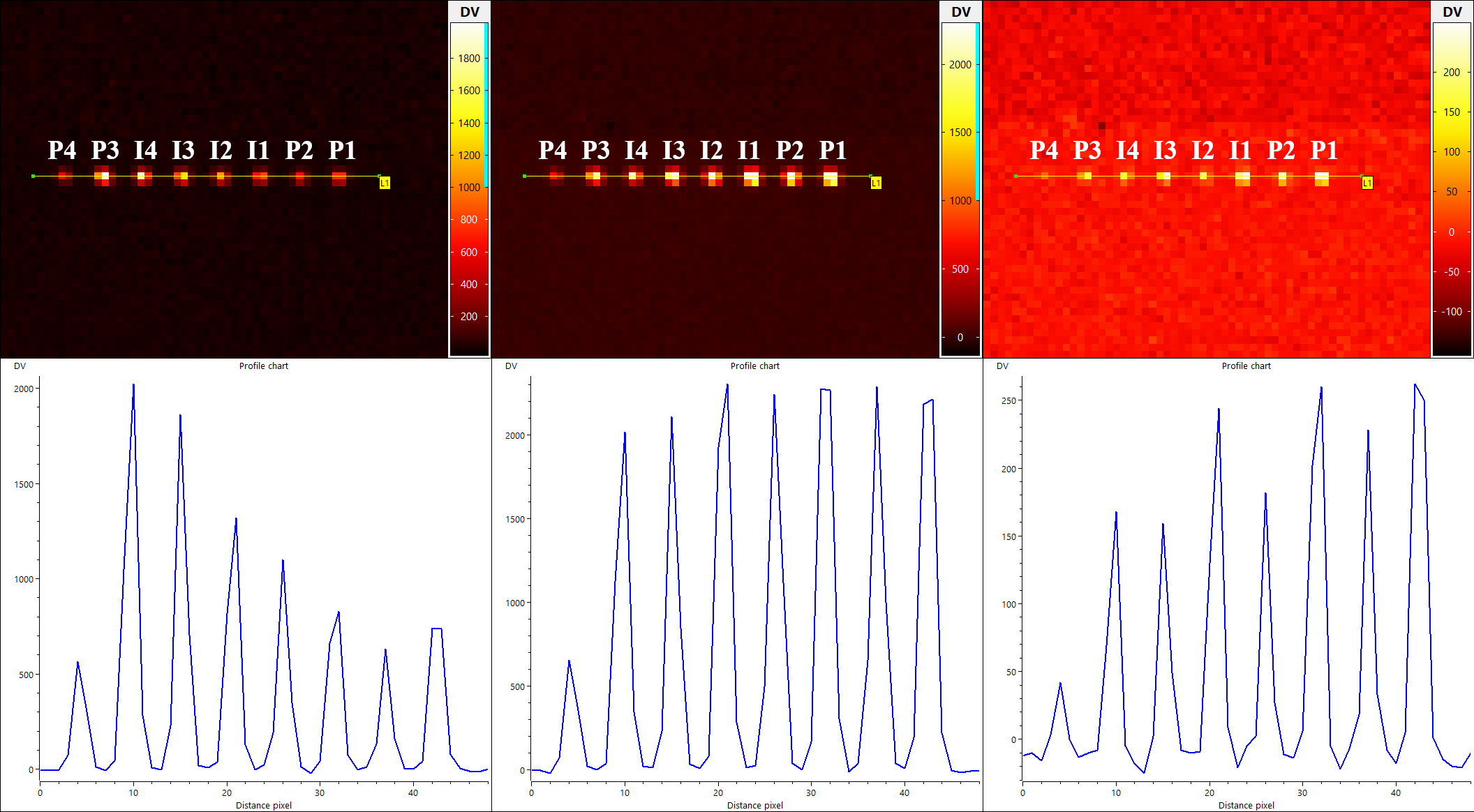}
        \caption{(top) Images of the chip's outputs after injection of the four beams in a nuller block. (bottom) Line profile of the detected flux. (left) Broadband injection with the blackbody source without optimization with the TTMs. (middle) Broadband injection with the blackbody source after optimization with the TTMs. (right) Injection with blackbody source and L'-band filter after optimization with the TTMs. All images are obtained with an integration time of 3.5\,ms.}
        \label{fig:IR camera results}
    \end{figure}

    Figure\,\ref{fig:lab status} shows the current status of the test bench assembly at KU Leuven. After the first assembling and alignment of the test bench following the procedures in Sec.\,\ref{sec:assembling & alignment strategy}, the light is injected inside the entrances of the photonic chip using the telecentric lens.
    This section summarizes the different experimental results obtained when injecting the light with the blackbody source. The four beams are injected simultaneously in a nuller block (described in Fig.\,\ref{fig:chip}) while the infrared camera records the eight generated outputs. Figure\,\ref{fig:IR camera results} confirms that even when using the blackbody source with an L'-band filter, the outputs are bright enough to be seen by the infrared camera.
    
\begin{figure}
        \centering
        \includegraphics[width=0.9\linewidth]{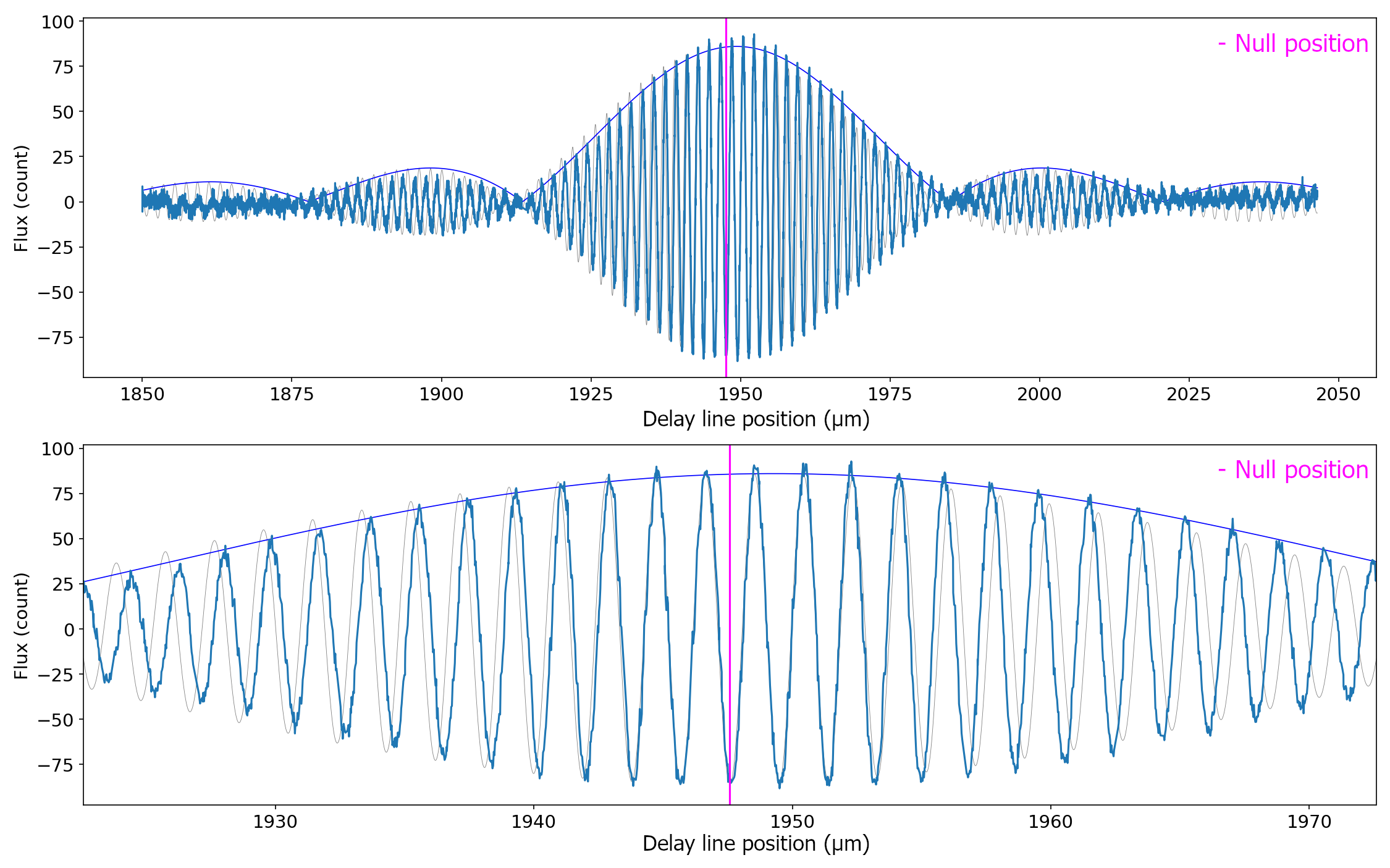}
        \caption{(top) Recorded fringes from output I1 during scanning of the delay line. The envelope of the fringe is observed (dark blue line). The position of the null is calculated from the fitting of the fringe packet and indicated by the pink vertical line. (bottom) Zoom of the top image around the null position.}
        \label{fig:find null}
    \end{figure}

\begin{figure}
        \centering
        \includegraphics[width = 0.49\linewidth]{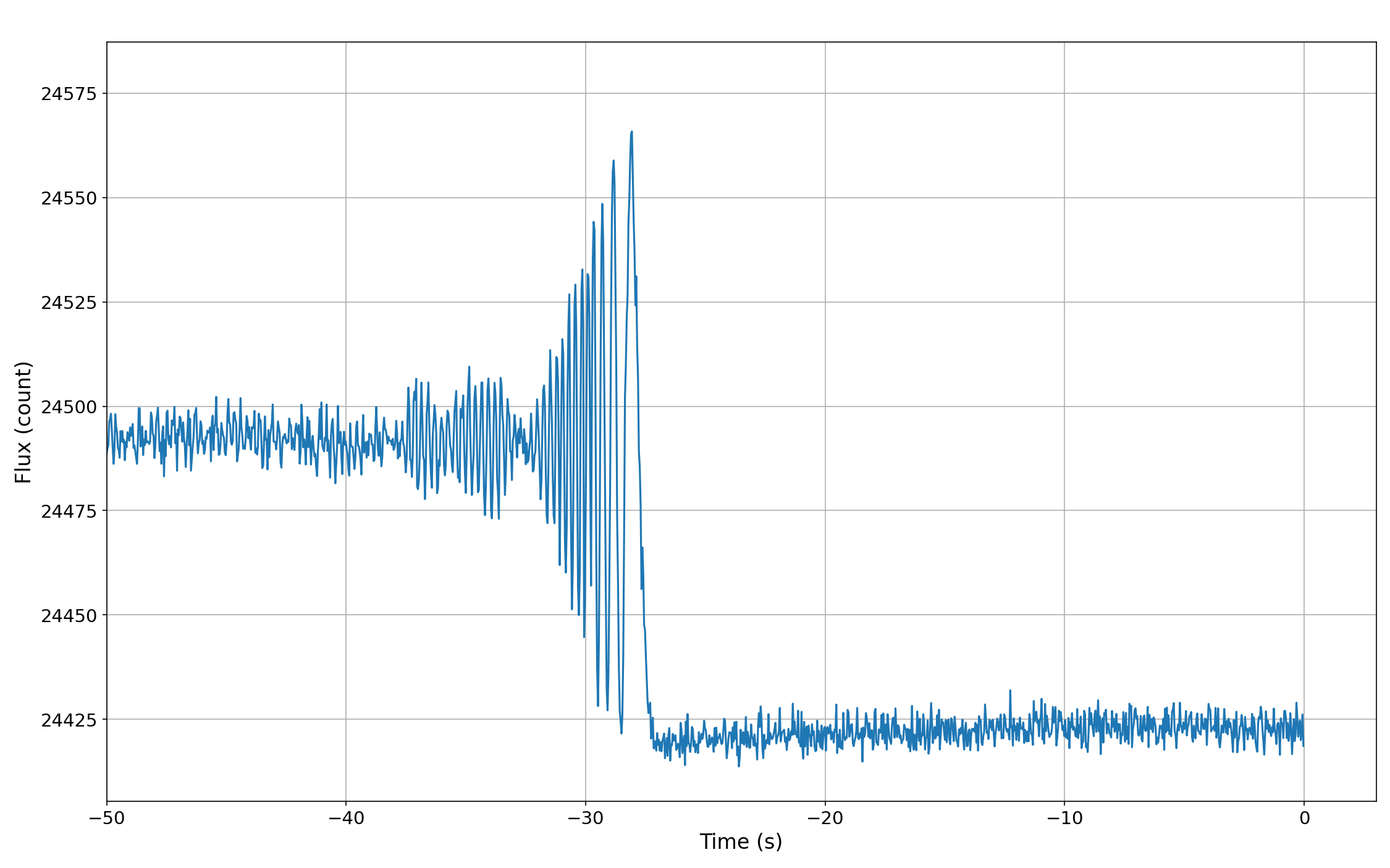}
        \includegraphics[width=0.49\linewidth]{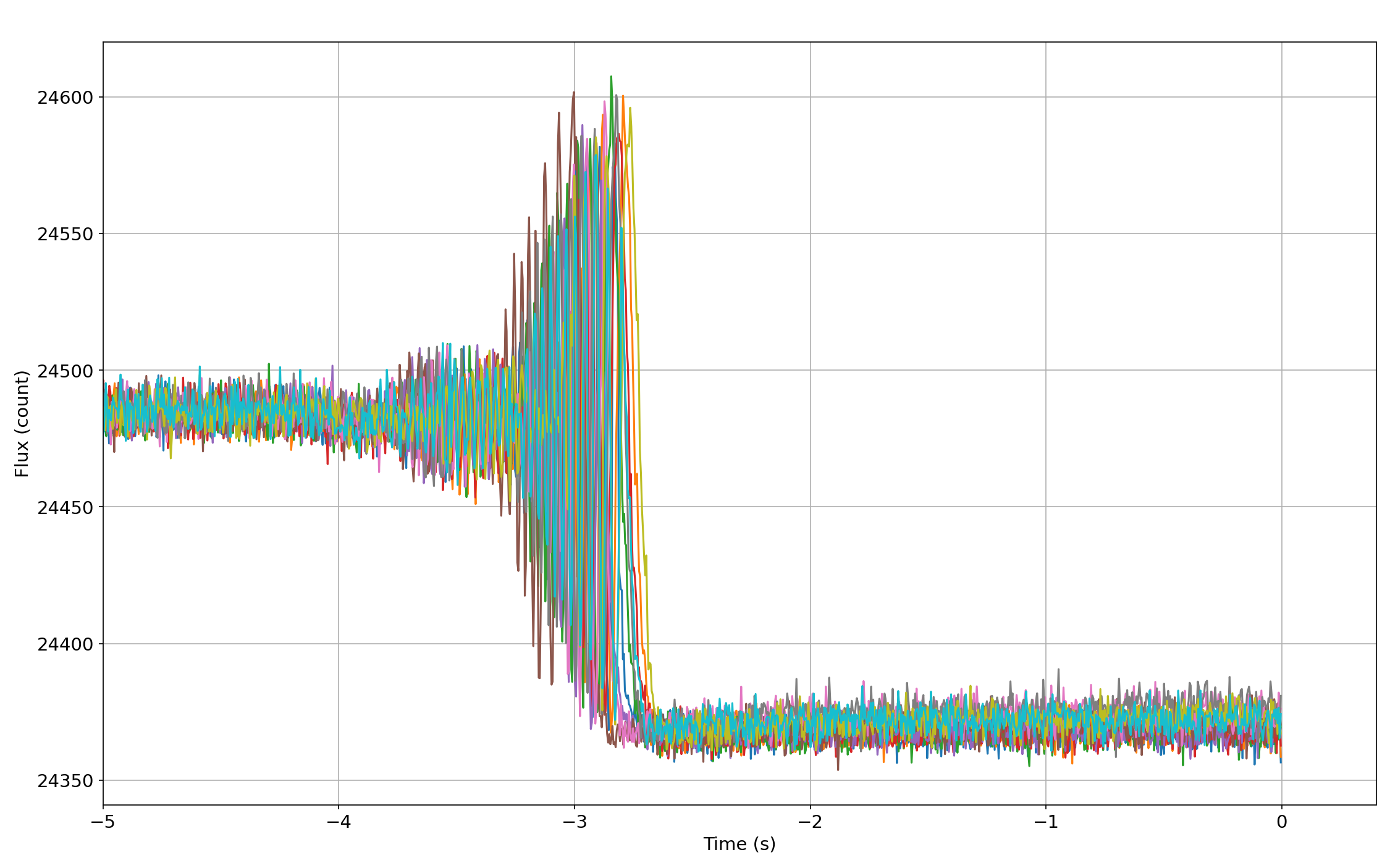}
        \caption{(left) Recorded flux from output I1 after finding the null position and feeding it to the delay line. The final flux corresponds to the minimum value of the fringe packet. This result shows that the delay line correctly stops around the null position. (right) Stack of 10 repeated recordings following the same procedure as the left image. This confirms the repeatability of the method to reach the null position.}
        \label{fig:reach null}
    \end{figure}

    After injection of the L'-band light and its optimization with the TTMs, the delay lines are used to scan the phase difference between the inputs 1 and 2 in Figure~\ref{fig:chip}. The infrared camera records the flux from output I1 during this scan. The obtained fringe packet is shown in Figure~\ref{fig:find null} along with its envelope. After fitting of the fringe packet, the null position where the inputs 1 and 2 are generating the deepest destructive interference is retrieved. When feeding this position to the delay line, the null can directly be reached with a sufficient repeatability (see Fig.\,\ref{fig:reach null}).

\section{CONCLUSION}
    The first lab assembly of the Asgard/NOTT warm optics is a benchmark for its integration within the Asgard instrument. Its main goals are threefold: (1) test and optimize the operation of all NOTT warm subsystems, (2) develop the software tools required to achieve deep interferometric nulls, (3) gain expertise and test the reliability of the assembling and alignment procedure. For this last goal, special attention was dedicated to the slicer which is an innovative solution to perform pupils recombination for infrared interferometry. 
    The design of the test bench is similar to the one used in Asgard, with the addition of two folding mirrors after the Pupils Recombination Optics for compactness on the optical table. This work is therefore also relevant for the future Asgard assembling in OCA and later in Paranal.
    The test bench was successfully assembled and aligned at KU Leuven using visible and infrared light. A monitoring system was used with visible light to help with the alignment and confirm the stability when switching between the green laser and the blackbody source.
    Using a blackbody source with an L'-band filter, the test bench was then used to inject light in a photonic chip similar to the one that will be used in Asgard/NOTT. The successful injection clearly showed the eight outputs of a nuller block, dedicated to performing nulling interferometry with the four beams. Using the delay lines, fringes were also obtained between two of the injected beams and the null position with the deepest destructive interference is retrieved. When feeding the null position to the delay line, the destructive interference is generated with sufficient repeatability. These results confirm the adequacy of the bench to test and optimize the Asgard/NOTT instrument at KU Leuven.

    
\section{FUTURE WORK}
    In the future, the hardware and software integration of the Asgard/NOTT warm sub-systems will continue, including the polarisation and dispersion control modules (Laugier et al. this conference). The automation of the system and acquisition software will also be improved to achieve deep interferometric nulls.
    The throughput of the test bench will be measured in the L'-band and compared to its estimation \citep{Garreau2024}. The characterization of the nuller blocks in the chip will also be performed and compared to the characterization performed in Köln (Sanny et al. this conference).
    An anti-reflection coating is envisioned for the surfaces of the chip to increase the throughput during the injection. To estimate the coating's efficiency, a characterization of the chip's throughput will be performed before and after applying the coating.
    Finally, a second characterization of the nuller blocks at cryo-temperature ($\sim150$\,K) will be carried out and compared to the Asgard/NOTT instrumental requirements. 
    Other technologies for integrated optics will also be characterized using this test bench, specifically lithium niobate photonic chips \citep{Heidmann2011,Martin2014,Martin2014b}. 

\subsection*{ACKNOWLEDGMENTS} 
SCIFY has received funding from the European Research Council (ERC); Award no. CoG –
866070 under the European Union’s Horizon 2020 research and innovation program. M-A.M. has
received funding from the European Union’s Horizon 2020 Research and Innovation Program;
Award No. 101004719.

\bibliographystyle{unsrtnat}
\bibliography{references}  






\end{document}